\documentclass[12pt]{article}

\usepackage[T1]{fontenc}
\usepackage{color}
\usepackage{float}
\usepackage{multirow}
\usepackage{amsmath,amssymb,amsthm,setspace,paralist}
\usepackage{graphicx}
\usepackage[authoryear]{natbib}
\usepackage{grffile}
\usepackage{lmodern}
\usepackage{algpseudocode}
\usepackage{algorithm}
\usepackage{algorithmicx}
\usepackage[colorlinks=true,linkcolor=blue, citecolor=blue,urlcolor=blue]{hyperref}

\makeatletter
\theoremstyle{plain}
\newtheorem{thm}{\protect\theoremname}

\providecommand{\lemmaname}{Lemma}
\providecommand{\theoremname}{Theorem}
\DeclareMathOperator*{\argmin}{arg\,min}
\DeclareMathOperator{\sgn}{sgn}

\@ifundefined{showcaptionsetup}{}{%
 \PassOptionsToPackage{caption=false}{subfig}}
 \usepackage{subfig}

\makeatother
\newcommand{\Breg}{{\mathbf{D}}}
\newcommand{\DKL}{\Breg_{\mbox{\tiny KL}}}

\begin{document}

\title{Iterative proportional scaling revisited: \\ a modern optimization perspective}

\author{Yiyuan She and Shao Tang\\Department of Statistics, Florida State University}
\date{}
\maketitle

{\small
\begin{abstract}
This paper revisits the classic iterative proportional scaling (IPS)  from a modern optimization perspective. In contrast
to the criticisms made in the literature, we show that based on a
coordinate descent characterization, IPS can be slightly modified
to deliver coefficient estimates, and from a majorization-minimization
standpoint, IPS can be extended to handle log-affine models with features not necessarily binary-valued or nonnegative. Furthermore,   some  state-of-the-art optimization techniques  such as block-wise computation, randomization and momentum-based acceleration can be employed to  provide  more scalable IPS algorithms, as well as some regularized variants  of IPS for concurrent feature selection.

\end{abstract}
}

\section{Introduction\label{sec:Introduction}}\label{sec:intro}
\subsection{Background}\label{subsec:background}
Count data are ubiquitous in modern statistical applications. Such
data are often cross-classified into contingency tables, where iterative
proportional scaling (IPS) can be applied as a standard tool \citep{Fienberg2006}.
IPS was firstly introduced by \citet{Deming1940} to adjust a contingency
table to obey prescribed column and row marginals, the problem of which
is  referred to as matrix raking nowadays. In general, IPS can be applied to Kullback\textendash Leibler
(KL) divergence minimization with linear constraints,
and Poisson log-linear model fitting on multi-way tables \citep{Ireland1968,Bishop1975}, and there is an interesting duality between these two types of problems
\citep{Good1963,Csiszar1975}. Theoretical studies regarding the
convergence properties of IPS undergo a long history and we refer
to \citet{Pukelsheim2014} for a comprehensive survey. To name a few,
\citet{Fienberg1970} took a geometric approach, assuming that all entries
in the input table are positive, while \citet{Haberman1974} and \citet{Bishop1975}
were among the first to use the ascent property of the associated
 log-likelihood. It is worth mentioning that the  theoretical studies become
more challenging if IPS operates on a table with zero entries \citep{sinkhorn1967b,Csiszar1975}.

Although the standard version of IPS is derived for contingency tables,
it has some quite useful and popular extensions. For example, \citet{Darroch1972} proposed
the generalized iterative scaling (GIS) which can fit Poisson log-affine
models with non-negative designs. Later, \citet{Pietra1997} developed
a more relaxed improved iterative scaling (IIS), without the requirement that all rows of the design matrix must sum to one. The class of IPS algorithms are widely used in the areas of Markov random fields and Gibbs distributions, natural language processing, matrix factorization, econometrics,  boosting,  and others
 \citep{sinkhorn1964,McCallum2000,Lahr2004,Phillips2004,kurras2015}. 

Today, IPS  is less frequently used by
statisticians,  largely due to the outstanding
performance and generality of Newton-type  algorithms.
In fact, Newton-Raphson is the default routine in most software for
fitting generalized linear models (GLMs), and can have quadratic convergence
in contrast to the linear convergence rate of IPS (ignoring the cost difference  per iteration).   Consider a five-way table of size $10\times10\times10\times10\times10$.
The associated Poisson model including all up to three-way interactions
has $8,\negthinspace146$ independent parameters. The corresponding
Hessian matrix has more than $10^{7}$ entries, leading to prohibitively
high computational cost when using Newton-Raphson. Quasi-Newton methods
are more computationally economical, but still fail easily when the
model is of high dimensionality. First-order methods (such as gradient
descent) are typically more scalable; in our problem, however, there
exists no universal stepsize due to the unbounded Hessian. This means
that some line search method must be adopted, but frequent function and gradient evaluations can be  expensive on big data.

The traditional  IPS algorithm is potentially helpful in this regard. 
In comparison with Newton-type and gradient descent methods, IPS has
certain benefits in computation. We illustrate the procedure via a
toy example. Consider a three-way table $m_{ijk}$
of size $m_{1}\times m_{2}\times m_{3}$ with the categorical variables
denoted by $X,Y,Z$ and assume a Poisson log-linear model
$(XY,YZ,XZ)$ with margins $m_{ij+}$, $m_{i+k}$
and $m_{+jk}$ as sufficient statistics  \citep{Agresti2012}. Starting with initial counts $[\mu_{ijk}^{(0)}]_{1\leq i\leq m_{1},1\leq j\leq m_{2},1\leq k\leq m_{3}}$,
IPS adjusts the estimated counts in a multiplicative manner iteratively:
$
\mu_{ijk}^{( t+1/3)}=\mu_{ijk}^{(t)}{m_{ij+}}/{\mu_{ij+}^{(t)}}$  for all $  (i,j)$, $\mu_{ijk}^{(t+2/3)}=\mu_{ijk}^{(t+1/3)} {m_{i+k}}/{\mu_{i+k}^{(t+1/3)}}$ for all $  (i,k)$, $\mu_{ijk}^{(t+1)}=\mu_{ijk}^{(t+2/3)} {m_{+jk}}/{\mu_{+jk}^{(t+2/3)}}
$ for all $  (j,k)$  
in  the $t$-th    epoch. Clearly, all calculations
  are ``in-place'' (with no need of
auxiliary variables) and no line search is required. The  procedure
has memory efficiency, scalability and implementation ease. In addition, it converges within one step once the solution has a closed form expression.

IPS is subject to some serious criticisms in the literature. Although IPS produces  expected cell counts, it does not deliver any coefficient estimate
nor asymptotic covariance estimate.  The elegant scaling procedure
  has however
a somewhat narrow scope and may encounter difficulties  say  in  the scenarios with features not necessarily bi-leveled or nonnegative,   or in  shrinkage estimation. Finally, though cost-effective
per iteration, IPS often requires a large number of iterations to converge.

This paper  attempts to  investigate IPS from a modern optimization perspective to {improve} and {generalize} the classic method and overcome the aforementioned
obstacles.  Our main contributions are threefold.
First, we are able to show that IPS implicitly involves a coefficient-update
step and adding it back leads to a novel coordinate descent  characterization of the procedure. To the best of our knowledge, this  is the first    fix of IPS  to produce  coefficient estimates.  Second, we reveal  an interesting connection of  IPS  to  majorization-minimization (MM) algorithms \citep{Hunter2004}. The MM principle   successfully generalizes  IPS to handle arbitrary features,
with the celebrated GIS and IIS taken as two particular
instances. Third,  we employ some  state-of-the-art optimization techniques,  such as block-wise computation, randomization and momentum-based acceleration,  to  develop  highly scalable IPS algorithms (without using parallel computation), as well as  some sparse variants of IPS for concurrent feature selection.

 The rest of the paper is organized as follows. Notations and model assumptions  are introduced in Section \ref{subsec:model}.  Section \ref{sec:IPSCD} describes a coefficient-driven IPS based on  coordinate descent, and  discusses its convergence properties,   efficient implementation and acceleration.  Section \ref{sec:A-Majorization-Minimization-View} shows several effective ways of constructing MM surrogate functions to generalize IPS.
Section \ref{sec:Shrinkage-estimation} develops sparse IPS  for high dimensional
estimation.     Section \ref{sec:exp}     shows the  accuracy and  efficiency    brought by  blockwise descent, randomization, reparametrization, and momentum-based acceleration  in simulations and real data experiments.

After the acceptance of our paper, we noticed the work by \citet{klimova2015}, which showed how to update the coefficients   in   IPS under the assumption of strictly positive counts.    In Section \ref{sec:IPSCD}  we give a more general result and more efficient algorithms. They also proposed two generalized  procedures   for  multinomial   likelihood optimization based on bisection or coarse-to-fine search.  See Section \ref{sec:A-Majorization-Minimization-View} for some fast     optimization algorithms     that can operate on general designs.

\subsection{Notation and model setting}
\label{subsec:model}
\textbf{Notation.} Given $N\in\mathbb{N}$, we define $[N]=\{1,2,\ldots,N\}$.
We use bold lower-case and upper-case symbols to denote column vectors
and matrices, respectively, i.e. $\boldsymbol{x}=[x_{i}]\in\mathbb{R}^{N}$
and $\boldsymbol{X}=[x_{ij}]\in\mathbb{R}^{N\times p}$, where $i\in[N]$
and $j\in[p]$. The inner product between two vectors $\boldsymbol{x}$
and $\boldsymbol{y}$ is   $\langle\boldsymbol{x},\boldsymbol{y}\rangle=\boldsymbol{x}^{T}\boldsymbol{y}$, their  Hadamard product is denoted by $\boldsymbol{x}\circ\boldsymbol{y}$, and their  component-wise division  is denoted by $\boldsymbol{x}\oslash\boldsymbol{y}$.
Given any matrix $\boldsymbol{X}$, we denote by $x_{i+}$ the $i$th
row sum $\sum_{j}x_{ij}$.
For notational ease, we extend all scalar functions and operations in a component-wise manner.  For example, given $\boldsymbol{x}=[x_{i}]\in\mathbb{R}^{N}$
and $\boldsymbol{y}\in\mathbb{R}^{N}$, $\exp(\boldsymbol{x})$ stands
for $[\exp(x_{i})]$, $\log(\boldsymbol x )= [\log x_i]$, and $\boldsymbol{x}\succeq\boldsymbol{y}$ indicates
that $x_{i}\geq y_{i}$ for all $i\in[N]$. The design matrix $\boldsymbol{X}\in\mathbb{R}^{N\times p}$
is frequently partitioned into columns (features) and rows: $\boldsymbol{X}=[\boldsymbol{x}_{1}\ldots\boldsymbol{x}_{p}]=[\tilde{\boldsymbol{x}}_{1}\ldots\tilde{\boldsymbol{x}}_{N}]^{T}$
with $\boldsymbol{x}_{j}\in\mathbb{R}^{N}$ and $\tilde{\boldsymbol{x}}_{i}\in\mathbb{R}^{p}$.
Given two square matrices $\boldsymbol{X}$ and $\boldsymbol{Y}$,
$\boldsymbol{X}\succeq\boldsymbol{Y}$ means $\boldsymbol{X}-\boldsymbol{Y}$
is positive semi-definite. We denote by $\lVert\boldsymbol{x}\rVert_{1}$
and $\lVert\boldsymbol{x}\rVert_{2}$ the $\ell_{1}$-norm and the
$\ell_{2}$-norm of $\boldsymbol{x}$, respectively. The spectral
norm and the infinity norm of a matrix $\boldsymbol{X}$ are defined
as $\lVert\boldsymbol{X}\rVert_{2}=\sigma_{\textrm{max}}(\boldsymbol{X})$
(the largest singular value of $\boldsymbol{X}$) and $\lVert\boldsymbol{X}\rVert_{\infty}=\max_{i}\sum_{j}\lvert x_{ij}\rvert$,
respectively. Given $\boldsymbol a, \boldsymbol b\in \mathbb R^N$ with $a_i, b_i\ge 0$, the (generalized) KL divergence is defined by $\DKL (\boldsymbol a \| \boldsymbol b) =  \sum_{i=1}^N [a_i \log (a_i/b_i) - a_i + b_i]$ and takes $+\infty$ when $a_i\ne 0$ and $b_i=0$ for some $i$.  The conventions
$\log0=-\infty$, $0\log(0/0)=0$, and $0\cdot(\pm\infty)=0$ are adopted.


Given a contingency table model, one can vectorize all cell counts and  introduce a design matrix  in the framework of Poisson  log-affine models. Concretely,
let $X_{k}$ ($1\leq k\leq r$) be the $k$th categorical variable
taking values in $[m_{k}]$ and $r$ be the total number of categorical
variables. Introduce dummy variables $X_{k}^{(\ell_{k})}\triangleq I(X_{k}=\ell_{k})$
for $\ell_{k}=2,3,\ldots,m_{k}$ with $X_k=\ell_1$ as the baseline, and let $\boldsymbol{X}_{k}=[X_{k}^{(2)},X_{k}^{(3)},\ldots,X_{k}^{(m_{k})}]$.
Then the model matrix containing all main effects has form $[\boldsymbol{1},\boldsymbol{X}_{1},\boldsymbol{X}_{2},\ldots,\boldsymbol{X}_{r}]$.
Similarly, the model matrix including all two-way interactions has
the following form $[\boldsymbol{X}_{1}*\!\boldsymbol{X}_{2},\ldots,\boldsymbol{X}_{1}*\!\boldsymbol{X}_{r},\boldsymbol{X}_{2}*\!\boldsymbol{X}_{3},\ldots,\boldsymbol{X}_{2}*\!\boldsymbol{X}_{r},\ldots,\boldsymbol{X}_{r-1}*\!\boldsymbol{X}_{r}]$,
where $\boldsymbol{X}_{j}*\!\boldsymbol{X}_{k}\triangleq[X_{j}^{(2)}X_{k}^{(2)},\ldots,X_{j}^{(2)}X_{k}^{(m_{k})},\ldots,X_{j}^{(m_{j})}X_{k}^{(m_{k})}]$.
Higher-order interactions can be included as well.
Given an arbitrary $\boldsymbol{X}\in\mathbb{R}^{N\times p}$, a
log-linear model with mean $\boldsymbol{\mu}\in\mathbb{R}^{N}$ satisfies   $\boldsymbol{\mu}=\exp(\boldsymbol{X}\boldsymbol{\beta}),$
where $\boldsymbol{\beta}$ denotes the coefficient vector. In some
applications, e.g., rate data analysis and matrix raking, an extra
offset $\boldsymbol{q}\succeq\boldsymbol{0}$ is required to specify
a \textit{log-affine} model  \citep{lauritzen1996}: $\boldsymbol{\mu}=\boldsymbol{q}\circ\exp(\boldsymbol{X}\boldsymbol{\beta})$ or $\mu_i = q_i \exp(\tilde {\boldsymbol x}_i^T \boldsymbol \beta)$.
We   allow $\beta_{j}$ to take $\pm\infty$, i.e., $\boldsymbol{\beta}\in\bar{\mathbb{R}}^{p}$
with $\bar{\mathbb{R}}=[-\infty,\infty]$ (the extended real line). Throughout the paper, we will not consider overdispersion or inflated zeros.

Let $\boldsymbol{n}\in\mathbb{R}^{N}$ with $n_i\ge 0$ denote the observed entries.
 The maximum likelihood (ML) estimation problem of $\boldsymbol{\mu}$
according to the log-affine model can be formulated by
\begin{equation}
\begin{aligned}\min_{\boldsymbol{\mu}}l(\boldsymbol{\mu}) & \triangleq-\langle\boldsymbol{n},\log\boldsymbol{\mu}\rangle+\langle\mathbf{1},\boldsymbol{\mu}\rangle\textrm{ s.t. }\boldsymbol{\mu}=\boldsymbol{q}\circ\exp(\boldsymbol{\eta}),\boldsymbol{\eta}\in\mathcal{\bar{R}}_{\boldsymbol{X}},\end{aligned}
\label{eq: opt MLE1}
\end{equation}
where $\mathcal{\bar{R}}_{\boldsymbol{X}}=\{\sum_{j=1}^{p}\beta_{j}\boldsymbol{x}_{j}\negmedspace:\negmedspace\beta_{j}\in\bar{\mathbb{R}}\}$
denotes the closure of the range of $\boldsymbol{X}$. Equivalently, the loss can  be  $\DKL(\boldsymbol n\|\boldsymbol \mu)$ or the deviance function. For convenience,
the constraint region is denoted by $\bar{\mathcal{M}}\triangleq\{\boldsymbol{\mu}\;\lvert\;\boldsymbol{\mu}=\boldsymbol{q}\circ\exp(\boldsymbol{\eta}),\boldsymbol{\eta}\in\mathcal{\bar{R}}_{\boldsymbol{X}}\}$.
As suggested by \citet{lauritzen1996}, theoretically it is more convenient  to
consider (\ref{eq: opt MLE1}) in a slightly more restricted manner
in order to guarantee the convergence of IPS:
\begin{equation}
\min_{\boldsymbol{\mu}} \DKL(\boldsymbol n\|\boldsymbol \mu)
\;\textrm{s.t. }\boldsymbol{\mu}\in\bar{\mathcal{M}}\cap\mathcal{M}^{*},\label{eq:opt problem MLE with space constraint}
\end{equation}
where $\mathcal{M}^{*}\triangleq\{\boldsymbol{\mu}\;\lvert\;l(\boldsymbol{\mu})<+\infty\}$,
meaning that $n_{i}>0$ implies $\mu_{i}>0$ ($1\leq i\leq N$).

For simplicity, the following assumptions are made throughout the
paper: (i) $\boldsymbol{n}\neq\boldsymbol{0}$, (ii) $\boldsymbol{x}_{j}\neq\boldsymbol{0},\;\forall j\in[p]$,
(iii) $\tilde{\boldsymbol{x}}_{i}\neq\boldsymbol{0},\;\forall i\in[N]$,
(iv) $\bar{\mathcal{M}}\cap\mathcal{M}^{*}\neq\emptyset$. Assumption
(i) is trivial. (ii) and (iii) are without loss of generality, since
one can drop the corresponding trivial predictors and/or observations.
(iv) ensures finite likelihood for at least one point in $\bar{\mathcal{M}}$,
and typically holds in real-life applications (otherwise one could add a mild $\ell_2$-type penalty to make the criterion strongly convex, cf. Section \ref{subsec:conv}).
It is worth mentioning that, under (iv), one can remove certain observations
to ensure a positive $\boldsymbol{q}$ without affecting the coefficient
estimation. In fact, for any $\boldsymbol{\mu}=[\mu_{i}]\in\bar{\mathcal{M}}\cap\mathcal{M}^{*}$,
$q_{i}=0$ implies $\mu_{i}=0$ and  $n_{i}=0$; so the $i$th
observation does not contribute to the objective function in (\ref{eq: opt MLE1})
and can be excluded in optimization. This  can greatly simplify
computation and analysis. 

To extend IPS, we specify three
types of design matrices:
\begin{equation}
(a)\:x_{ij}=0\:\textrm{or}\:1\:(\textbf{binary}),\:(b)\:x_{ij}\geq0\:(\textbf{non-negative}),\:(c)\:x_{ij}\in\mathbb{R}\:(\textbf{general}).\label{eq:designs}
\end{equation}
Clearly,   design matrices derived from contingency table models
are special cases of (a).  But real applications can go much beyond this binary setting.

\section{A Coordinate Descent Characterization}
\label{sec:IPSCD}

As mentioned in Section \ref{sec:intro}, IPS is commonly used for matrix raking  to     find  a table $\boldsymbol{\mu}$ that not only matches the marginals of a reference table $\boldsymbol n\in \mathbb R^N$ but is closest to an initial (prior) $\boldsymbol q\in \mathbb R^N$ in the sense of   relative entropy or KL divergence:
\begin{align}
\min_{\boldsymbol \mu\in \mathbb R^N} \DKL (\boldsymbol \mu\|\boldsymbol q) =  \sum_{i=1}^N [\mu_i \log (\mu_i/q_i) - \mu_i + q_i], \mbox{ s.t. } \langle {\boldsymbol{x}}_j, \boldsymbol \mu\rangle=\langle {\boldsymbol{x}}_j, \boldsymbol n\rangle, j\in [p].\label{probKL}
\end{align}
The features ${\boldsymbol x}_j$ are binary-valued in the table setup (but not so in general) and $\langle {\boldsymbol{x}}_j, \boldsymbol n\rangle$ give  prescribed marginals. In certain applications there is no need to provide $\boldsymbol n$, since only the constraint values are  needed.   When $\boldsymbol 1 \in \mathcal R(\boldsymbol X)$, $\langle \boldsymbol 1, \boldsymbol \mu\rangle$ is a constant and so the objective can be reduced to $\sum \mu_i \log (\mu_i/q_i)$. Without the binary restriction on the design,  \eqref{probKL} defines the general problem of   \textit{maximum entropy} with linear constraints, and has widespread applications in   statistical mechanics, information theory,  natural language processing, and ecology \citep{Berger1996,Dudik2004,Elith2011}.
The duality between the maximum entropy problem \eqref{probKL} and the maximum likelihood  problem \eqref{eq: opt MLE1} is well known by statisticians, and indeed many take the ML route   to study the  properties of IPS. 

\subsection{Coefficient-driven IPS}
\label{subsec:coeffIPS}
IPS is often criticized for not being able to deliver a coefficient
estimate. We will show, however, that this is not true and IPS includes
an implicit coefficient-update step.

First, although the objective function of IPS is conventionally formulated
with respect to the unknown mean vector
  $\boldsymbol{\mu}$ (or $\log\boldsymbol{\mu}$) in the literature,   it is arguably more insightful to rewrite (\ref{eq: opt MLE1})
in terms of $\boldsymbol{\beta}$:
\begin{equation}
\min_{\boldsymbol{\beta\in \bar{\mathbb{R}}^{p}}}l(\boldsymbol{\beta})=-\langle\boldsymbol{n},\boldsymbol{X}\boldsymbol{\beta}\rangle+\langle\boldsymbol{q},\exp(\boldsymbol{X}\boldsymbol{\beta})\rangle,\label{eq:opt problem MLE beta}
\end{equation}
where we used $\mu_i(\boldsymbol \beta)=q_i\exp(\tilde{\boldsymbol x}_i^T \boldsymbol \beta)$.  The convenience can be partially observed from the initialization
condition of IPS \citep{Fienberg2006}, which requires $\boldsymbol{\mu}^{(0)}$ to take the form
of $\boldsymbol{q}\circ\exp(\boldsymbol{\eta}^{(0)})$ for some $\boldsymbol{\eta}^{(0)}$
in $\mathcal{{R}}_{\boldsymbol{X}}$.   Here,   we still use $l(\cdot)$ to denote the loss by abuse of notation, and when   $\boldsymbol 1 \in \mathcal R(\boldsymbol X)$, it is easy to see that \eqref{eq:opt problem MLE beta} amounts to minimizing   the $G^2$-statistic   $2\sum n_i \log (n_i/\mu_i(\boldsymbol \beta))$. For
this convex optimization problem, we can design a simple \textit{cyclic coordinate
descent} (CD) algorithm, which updates $\beta_{j}$ ($1\le j\le p$) according to the following formula with the other
coordinates fixed
\begin{equation}
\beta_{j}^{(t+1)}\in\textrm{arg}\min_{\beta_{j}}l(\beta_{1}^{(t+1)},\ldots,\beta_{j-1}^{(t+1)},\beta_{j},\beta_{j+1}^{(t)},\ldots,\beta_{p}^{(t)}).\label{eq:cd univarite problem}
\end{equation}
Define $\boldsymbol{\mu}^{(t,j-1)}\triangleq\boldsymbol{q}\circ\exp(\boldsymbol{X}\boldsymbol{\beta}^{(t,j-1)})$
with $\boldsymbol{\beta}^{(t,j-1)}\triangleq[\beta_{1}^{(t+1)},\ldots,\beta_{j-1}^{(t+1)},\beta_{j}^{(t)},\allowbreak\beta_{j+1}^{(t)},\ldots,\beta_{p}^{(t)}]^{T}$, $j\in [p]$.
It is easy to show that $\beta_{j}^{(t+1)}$ satisfies the equation \begin{equation}
(\boldsymbol{\mu}^{(t,j-1)})^{T}\{\boldsymbol{x}_{j}\circ\exp[\boldsymbol{x}_{j}(\beta_{j}-\beta_{j}^{(t)})]\}-\boldsymbol{x}_{j}^{T}\boldsymbol{n}=0.\label{eq:cd optimal}
\end{equation}
Algorithm \ref{alg:CD alg} shows the details of the cyclic coordinate
update.
 The solution to (\ref{eq:cd optimal}) has a closed form in some
  cases. For example, if we assume the design is derived from
a contingency table model, or more generally, $\boldsymbol{X}$ is
binary (cf. (\ref{eq:designs})), then the   following crucial fact
\begin{equation}
x_{ij}\exp(x_{ij} \beta_{j} )=x_{ij}\exp(\beta_{j})\label{eq:binary}
\end{equation}
implies  that $\boldsymbol{x}_{j}\circ\exp[(\beta_{j}-\beta_{j}^{(t)})\boldsymbol{x}_{j}]=\boldsymbol{x}_{j}\exp(\beta_{j}-\beta_{j}^{(t)})$,
and
so\begin{equation}
\beta_{j}^{(t+1)}=\beta_{j}^{(t)}+\log[\boldsymbol{x}_{j}^{T}\boldsymbol{n}/(\boldsymbol{x}_{j}^{T}\boldsymbol{\mu}^{(t,j-1)})].\label{eq:analytical solution cd}
\end{equation}
Step 5 in Algorithm \ref{alg:CD alg} then becomes
\begin{equation}
\boldsymbol{\mu}^{(t,j)}=\boldsymbol{\mu}^{(t,j-1)}\circ\exp\{\boldsymbol{x}_{j}\log[(\boldsymbol{x}_{j}^{T}\boldsymbol{n})/(\boldsymbol{x}_{j}^{T}\boldsymbol{\mu}^{(t,j-1)})]\},\label{eq:cd mu}
\end{equation}
or equivalently, $\mu_{i}^{(t,j)}=(\langle\boldsymbol{x}_{j},\boldsymbol{n}\rangle/\langle\boldsymbol{x}_{j},\boldsymbol{\mu}^{(t,j-1)}\rangle)\mu_{i}^{(t,j-1)}$
if $x_{ij}=1$, and $\mu_{i}^{(t,j)}=\mu_{i}^{(t,j-1)}$ otherwise
($1\leq i\leq N$), which is exactly the IPS algorithm used in matrix raking. In the literature,
\citet{Haberman1974} shows that IPS is a cyclic ascent method in
updating $\log\boldsymbol{\mu}$, but to the best of our knowledge, formulating
IPS as cyclic CD on $\boldsymbol{\beta}$ is new.

{\begin{algorithm}[t!] \small\caption{\textbf{IPS-CD} \label{alg:CD alg}}         \textbf{Input} $\boldsymbol{n}$, $\boldsymbol{q}$ and $\boldsymbol{X}$\\         \textbf{Initialize} $\boldsymbol{\beta}^{(0)}\in\mathbb{R}^{p}$, $t\leftarrow0$         \begin{algorithmic}[1]
\While{  not converged }
\State                 $\boldsymbol{\mu}^{(t,0)}\leftarrow\boldsymbol{q}\circ\exp(\boldsymbol{X}\boldsymbol{\beta}^{(t)})$
\For {$j=1,2,\cdots,p$}
\State  $\beta_{j}^{(t+1)}\in\argmin_{\beta_{j}}l(\beta_{1}^{(t+1)},\ldots,\beta_{j-1}^{(t+1)},\beta_{j},\beta_{j+1}^{(t)},\ldots,\beta_{p}^{(t)})$. Binary case:\ cf. (\ref{eq:analytical solution cd})
\State $\boldsymbol{\mu}^{(t,j)}\leftarrow\boldsymbol{\mu}^{(t,j-1)}\circ\exp[\boldsymbol{x}_{j}(\beta_{j}^{(t+1)}-\beta_{j}^{(t)})]$. Binary case: cf. (\ref{eq:cd mu})
   \EndFor
\State
$\boldsymbol{\beta}^{(t+1)}\leftarrow[\beta_{1}^{(t+1)},\ldots,\beta_{p}^{(t+1)}]^{T}$,  $t\leftarrow t+1$                 \EndWhile  \\

\Return
$\hat{\boldsymbol{\mu}}=\boldsymbol{\mu}^{(t-1,p)}$, $\hat{\boldsymbol{\beta}}=\boldsymbol{\beta}^{(t)}$         \end{algorithmic} \end{algorithm}}

\begin{figure}[h!]
\begin{centering}
\subfloat[Traditional IPS]{\centering{}\includegraphics[width=0.5\columnwidth]{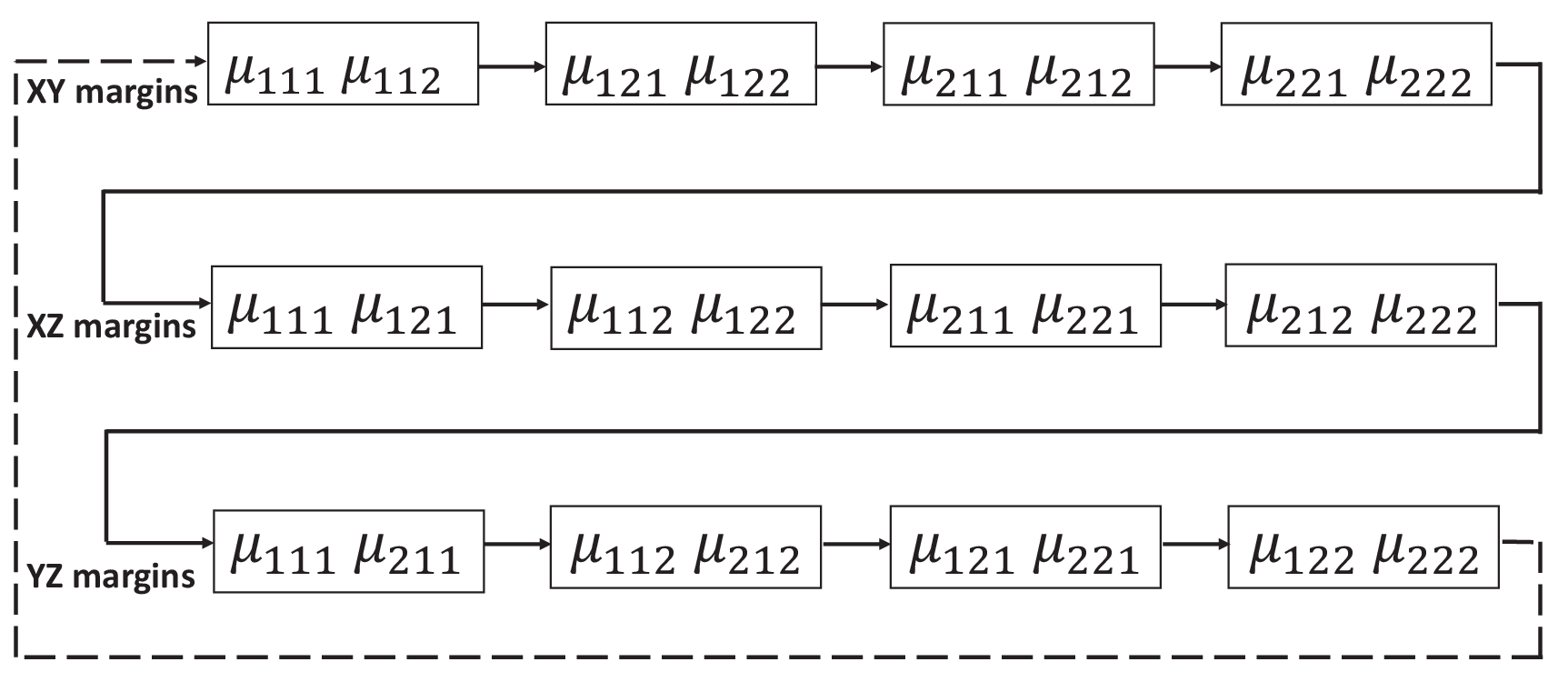}}
\subfloat[IPS-CD ]{\centering{}\includegraphics[width=0.5\columnwidth]{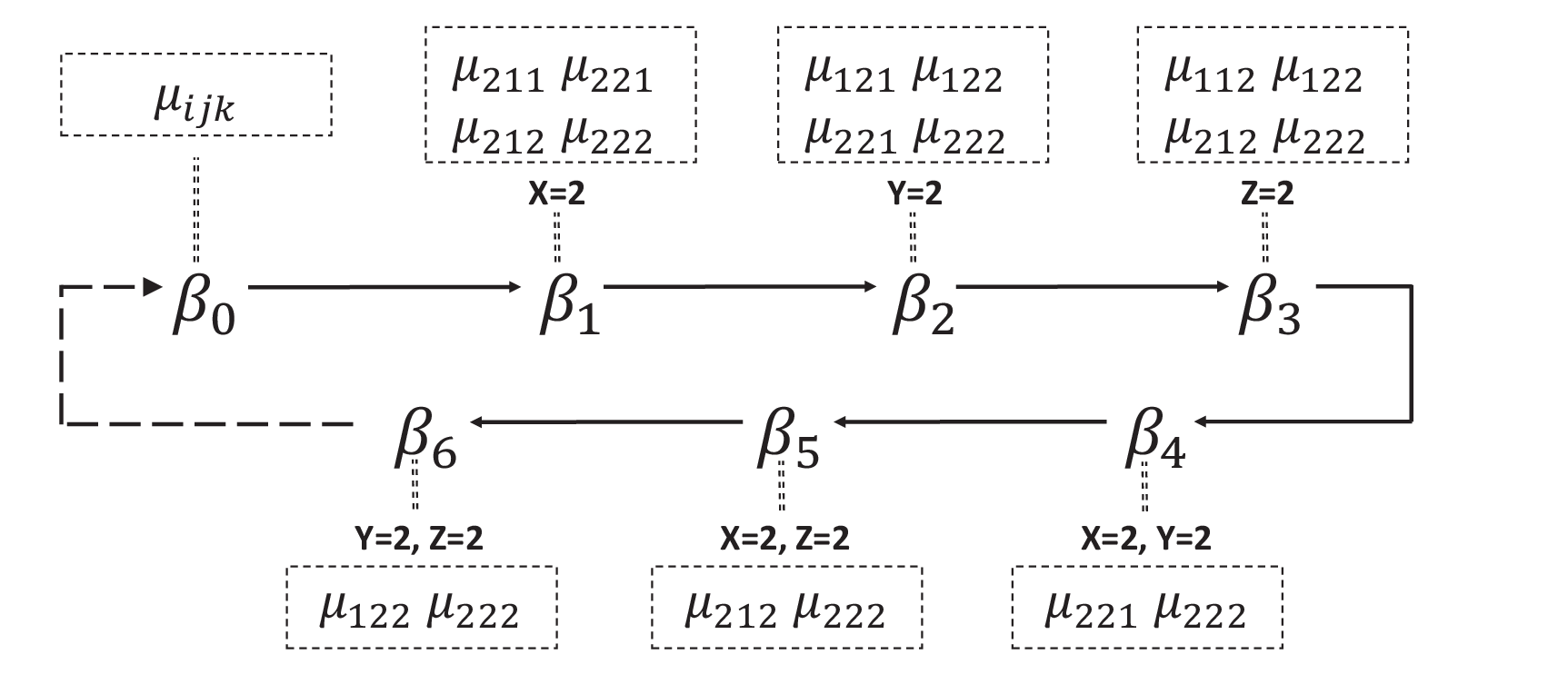}}\caption{\footnotesize Update-scheme comparison on a $2\times2\times2$ three-way table with $X, Y, Z$ taking  levels $1, 2$. The homogeneous
association model $(XY,XZ,YZ)$ is assumed, with $XY,XZ,YZ$ margins as  minimal
sufficient statistics. Here,   $\beta_{0}$ is the intercept and $\beta_{1}$,
$\beta_{2}$, $\beta_{3}$, $\beta_{4}$,
$\beta_{5}$, $\beta_{6}$ denote the coefficients of $I({X=2})$, $I({Y=2})$, $I({Z=2})$, $I({X=2,Y=2})$, $I({X=2,Z=2})$, and $I({Y=2,Z=2})$, respectively. In each epoch, (a) updates the cell values  in 12 steps to satisfy prescribed margins, while (b) updates $\beta$ and the associated $\mu_{ijk}$ in 7 steps. \label{fig:ipscdscheme}}
\end{centering}
\end{figure}

Algorithm \ref{alg:CD alg} provides more flexibility
in initialization. For instance, $\boldsymbol{\mu}^{(0)}=\boldsymbol{q}$
is unnecessary; rather, starting with an arbitrary $\boldsymbol{\beta}^{(0)}\in\mathbb{R}^{p}$
suffices. In our experience, a properly chosen initial point can reduce
the computational time substantially in large-scale data problems.
The algorithm not only offers an easy fix of IPS to yield $\hat{\boldsymbol{\beta}}$,
but also suggests efficient ways to update multiple components of
$\boldsymbol{\mu}$ at one time. Concretely, when $\beta_{j}$ ($1\le j\le p$) is changed, all the $\mu_{i}$ with
$x_{ij}\neq0$ can be updated. It is thus  extremely helpful in implementation to consolidate the features  and use a design matrix with full column rank (which can be easily obtained by QR or LU  decomposition). For example, on the aforementioned homogeneous
association model $(XY,XZ,YZ)$    with $X, Y, Z$ taking   two levels,  each epoch of the ordinary IPS  updates all cell values according to the   minimal
sufficient statistics  in $4\times 3=12$ steps (cf. Section \ref{subsec:background}), while IPS-CD updates $\beta$ and the associated cells in 7 steps; see Figure \ref{fig:ipscdscheme}
for an illustration.

Some people conjectured that IPS    attempted to minimize Pearson's $X^{2}$-statistic,
but this is not true---see  Theorem  \ref{cor:G^2}. It is well known that $X^{2}$ converges
 faster than $G^{2}$ to the asymptotic $\chi^{2}$-distribution. Nicely,
our coordinate descent standpoint can  modify  IPS
in a simple way to solve the problem $\min_{\boldsymbol{\mu}}\sum_{i}[n_{i}-\mu_{i}(\boldsymbol{\beta})]^{2}/\mu_{i}(\boldsymbol{\beta})$
with $\mu_{i}(\boldsymbol{\beta})=q_{i}\exp(\tilde{\boldsymbol{x}}_{i}^{T}\boldsymbol{\beta})$.
In fact, similar to the derivation of (\ref{eq:cd optimal}), given the other
coordinates $\beta_{j}^{(t+1)}$ can be updated by solving the equation
\begin{gather*}
\sum_{i}x_{ij}\mu_{i}^{(t,j-1)}\exp[x_{ij}(\beta_{j}-\beta_{j}^{(t)})]-\sum_{i}{x_{ij}n_{i}^{2}}/\{{\mu_{i}^{(t,j-1)}\exp[x_{ij}(\beta_{j}-\beta_{j}^{(t)})]}\}=0,
\end{gather*} and 
 with a binary design, the equation has a closed-form
solution and the resultant iterative  scaling of $\boldsymbol{\mu}$
is
\begin{gather*}
\boldsymbol{\mu}^{(t,j)}=\boldsymbol{\mu}^{(t,j-1)}\circ\exp\{\frac{1}{2}\boldsymbol{x}_{j}\log[\boldsymbol{x}_{j}^{T}(\boldsymbol{n}\circ\boldsymbol{n}\oslash\boldsymbol{\mu}^{(t,j-1)})/(\boldsymbol{x}_{j}^{T}\boldsymbol{\mu}^{(t,j-1)})]\}.
\end{gather*}
The procedure can be conveniently used for testing associations and interactions in contingency tables.
Its differences from the ordinary version  (\ref{eq:cd mu}) are seen in the term  $\boldsymbol{n}\circ\boldsymbol{n}\oslash\boldsymbol{\mu}^{(t,j-1)}$ in place of $\boldsymbol{n}$, and the
additional factor of $1/2$.

\subsection{Convergence properties}
\label{subsec:conv}
The CD characterization facilitates  theoretical
studies of the convergence properties of IPS. First, we have a
natural outcome for Algorithm \ref{alg:CD alg} for a general design
$\boldsymbol{X}$.
\begin{thm}
\label{cor:G^2}For the sequence of iterates $\{\boldsymbol{\beta}^{(t)}\}_{t=0}^{\infty}$
generated by Algorithm \ref{alg:CD alg}, the associated function
values $l(\boldsymbol{\beta}^{(t)})$ are monotonically non-increasing
for $t\geq0$. In particular, if the first column of $\boldsymbol{X}$
corresponds to the intercept, then the $G^{2}$-statistic evaluated
on $\boldsymbol{\mu}^{(t,1)}$, i.e., $2\sum_{i}n_{i}\log(n_{i}/\mu_{i}^{(t,1)})$,
is monotonically non-increasing.
\end{thm}
Similar results have been obtained on contingency tables   (e.g., \cite{Bishop1975}).
The theorem  directly follows from the coordinate
descent nature of the algorithm design, and is not restricted to
tables.
Moreover, because of the convexity of the problem, under some regularity
conditions, the convergence of the sequence of iterates is readily
at hand.
\begin{thm}
\label{prop:sequence convergence CD}Suppose that there exists a unique
solution $\hat{\boldsymbol{\beta}}$ to problem (\ref{eq:opt problem MLE beta})
of finite norm. Then the sequence of iterates $\{\boldsymbol{\beta}^{(t)}\}_{t=0}^{\infty}$
generated by Algorithm \ref{alg:CD alg} has a unique limit point
$\hat{\boldsymbol{\beta}}$, and the rate of convergence is at least
linear.
\end{thm}
The linear convergence rate was  shown by \citet{Fienberg1970} for matrix raking. Our theorem builds upon the theory of descent methods \citep{Luo1992}.
In consideration of recent advances \citep{Tseng2001,Razaviyayn2013},
one can possibly relax the assumption of Theorem \ref{prop:sequence convergence CD} in certain ways, but we will
not pursue further in this work. We refer to \cite{Fienberg2012} and \cite{wang2016}
for detailed studies of the existence and uniqueness of MLE. In practice,
we could add a mild ridge-type penalty (cf. Section \ref{sec:Shrinkage-estimation}),
which ensures the condition and  enhances numerical stability.

\subsection{Randomization and block-wise computation \label{sub: Randomization}}
Recently, CD algorithms have received a lot of attention  in  statistics (high dimensional statistics, in particular)
due to their simplicity and low-complexity operations at each iteration.
But the         characteristic of  not updating all variables together may also make them     take more iterations and require more stringent conditions to converge.
Instead of the cyclic update in Algorithm
\ref{alg:CD alg},   one can choose the coordinate
with the  largest   derivative in magnitude, $j=\textrm{arg}\max_{j}\lvert\nabla_{j}l(\boldsymbol{\beta}^{(t)})\rvert$,
the so-called Gauss-Southwell (G-S) rule. G-S can      successfully reduce the number of iterations, and has been    analyzed in depth in the literature (see, e.g., \cite{Bertsekas2015} and \cite{nutini2015}).
However, it is  inefficient in large problems since all partial derivatives have to be computed---indeed, with the full gradient
vector available, one could update the whole vector $\boldsymbol{\beta}$
at each step.

An effective way to speed the convergence of the algorithm is to \textit{randomize} it. The recent analysis of \cite{Nesterov2012}  shows that random coordinate selection
 can achieve the same convergence rate as G-S.  It is worth mentioning that  random strategies   make complexity bounds much easier to obtain, and are often  suitable for modern computational architectures. Experience shows that   random sampling
without replacement \citep{wright2015}  works well in IPS.  Concretely, at the start
of each cycle, we  randomly shuffle the elements in $[p]$ to obtain
a new index set $\mathrm{Perm}[p]$, and then update the permuted
coordinates sequentially.  The randomized algorithm, denoted by \textbf{A}ccelerated-\textbf{IPS}
(\textbf{A-IPS}), is presented in Algorithm \ref{alg:AIPS}.
In theory, randomized   coordinate selection is able to avoid worst-case order of coordinates, and seems to be better
than the cyclic rule in an average sense \citep{Richtarik2014}.  (Sampling the coordinates in a data-dependent manner  may be better, but it involves more computation and  is not considered here.)
{\begin{algorithm}[t!] \small\caption{\textbf{A-IPS}.\label{alg:AIPS}}           \textbf{Input} $\boldsymbol{n}$, $\boldsymbol{q}$ and $\boldsymbol{X}$\\         \textbf{Initialize} $\boldsymbol{\beta}^{(0)}\in\mathbb{R}^{p}$, $t\leftarrow0$         \begin{algorithmic}[1]
\While{  not converged }
\State                 $\boldsymbol{\mu}^{(t,0)}\leftarrow\boldsymbol{q}\circ\exp(\boldsymbol{X}\boldsymbol{\beta}^{(t)})$
\For {$j$ \textbf{in $\mathrm{Perm}[p]$}}
\State  $\beta_{j}^{(t+1)}\in\argmin_{\beta_{j}}l(\beta_{1}^{(t+1)},\ldots,\beta_{j-1}^{(t+1)},\beta_{j},\beta_{j+1}^{(t)},\ldots,\beta_{p}^{(t)})$
\State $\boldsymbol{\mu}^{(t,j)}\leftarrow\boldsymbol{\mu}^{(t,j-1)}\circ\exp[\boldsymbol{x}_{j}(\beta_{j}^{(t+1)}-\beta_{j}^{(t)})]$
\EndFor
\State
$\boldsymbol{\beta}^{(t+1)}\leftarrow[\beta_{1}^{(t+1)},\ldots,\beta_{p}^{(t+1)}]^{T}$,  $t\leftarrow t+1$                 \EndWhile \\                 \Return $\hat{\boldsymbol{\mu}}=\boldsymbol{\mu}^{(t-1,p)}$, $\hat{\boldsymbol{\beta}}=\boldsymbol{\beta}^{(t)}$         \end{algorithmic} \end{algorithm}}{\small \par}

\begin{figure}
\begin{centering}
\includegraphics[width=.8\columnwidth]{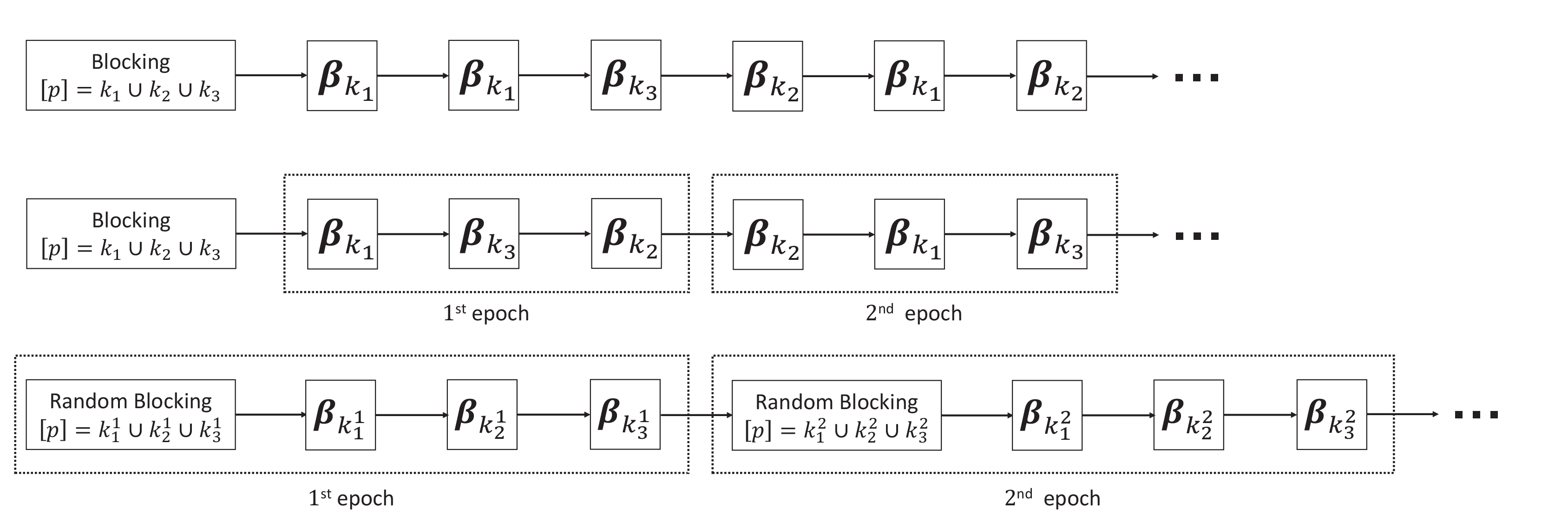}
\caption{\footnotesize Update schemes of block resampling with replacement (top), block
resampling without replacement (middle), and random blocking followed by cyclic update (bottom) which is the recommended approach. \label{fig-blockupdates}} 
\end{centering}
\end{figure}

Extending the coordinate-by-coordinate update to block-by-block update may  save some computational time, too.    Block coordinate descent (BCD)   decomposes all  unknowns into $m$ blocks, and  updates only one  block at a time.
In this way,
a large difficult problem can be reduced to a series of smaller and easier sub-problems
  so that Newton or quasi-Newton methods  such as L-BFGS   \citep{liu1989}   can be applied
with ease. In the block setting,  again,
G-S  takes much fewer iteration steps than the plain cyclic
rule, but in terms of  total computational time,  it incurs
significantly more overhead in big data applications due to the calculation
of the full gradient. Perhaps surprisingly, we found that some
well-known random rules, such as (block) resampling without replacement,
may not  have satisfactory performance, either. We recommend
\textit{random blocking} followed by \textit{cyclic update}, and this   random block-wise IPS is referred to as   \textbf{B-IPS}.  See Figure \ref{fig-blockupdates} for an illustration. At the beginning
of each iteration, all coordinates are randomly shuffled and
sequentially assigned into $m$ blocks with sizes $g_{1},\ldots,g_{m}$, and then we update the blocks of coefficients in a cyclic manner.  This  strategy  substantially outperforms  the other  rules according to our experiments.

\section{A Majorization-Minimization Viewpoint \label{sec:A-Majorization-Minimization-View}}

This section uses the majorization-minimization (MM) principle
to study and generalize IPS. We refer to \citet{lange2000} and \citet{Hunter2004}
for more details on MM algorithms. Rather than directly minimizing
$l(\boldsymbol{\beta})$ in (\ref{eq:opt problem MLE beta}), our
goal here is to construct a surrogate function $g(\boldsymbol{\beta}\,|\,\boldsymbol{\beta}^{-})$,
with $\boldsymbol{\beta}^{-}$ denoting the value of $\boldsymbol{\beta}$
from the last iteration, such that the following properties are satisfied
for all $\boldsymbol{\beta}\in\bar{\mathbb{R}}^{p}$:
\begin{equation}
g(\boldsymbol{\beta}\mid\boldsymbol{\beta}^{-})\geq l(\boldsymbol{\beta})\;,\;g(\boldsymbol{\beta}\mid\boldsymbol{\beta})=l(\boldsymbol{\beta}).\label{eq:MM property}
\end{equation}
Then, if we set $\boldsymbol{\beta}^{(t+1)}\in\textrm{arg}\min_{\boldsymbol{\beta}}g(\boldsymbol{\beta}\mid\boldsymbol{\beta}^{(t)})$,
the sequence of iterates satisfies
\[
l(\boldsymbol{\beta}^{(t+1)})\leq g(\boldsymbol{\beta}^{(t+1)}\mid\boldsymbol{\beta}^{(t)})\leq g(\boldsymbol{\beta}^{(t)}\mid\boldsymbol{\beta}^{(t)})=l(\boldsymbol{\beta}^{(t)}).
\]
Hence   repeatedly minimizing the surrogate function $g(\cdot\,|\,\boldsymbol{\beta}^{-})$
 guarantees the original objective function $l(\boldsymbol{\beta})$
to be monotonically non-increasing. Because the problem under consideration is convex, there are other nice properties
of the   MM iterates  (e.g., Chapter 12 of \citet{lange2013}). We shall focus on the derivation
of surrogate functions in this section. For notational simplicity,
 define $\boldsymbol{\mu}^{-}\triangleq\exp(\boldsymbol{X}\boldsymbol{\beta}^{-})$
and $\boldsymbol{\mu}^{(t)}\triangleq\exp(\boldsymbol{X}\boldsymbol{\beta}^{(t)})$.
After getting $\boldsymbol{\beta}^{(t+1)}$, one can update $\boldsymbol{\mu}$
by
\begin{equation}
\boldsymbol{\mu}^{(t+1)}=\boldsymbol{\mu}^{(t)}\circ\exp[\sum_{j}\boldsymbol{x}_{j}(\beta_{j}^{(t+1)}-\beta_{j}^{(t)})],\label{eq:MM mu}
\end{equation}
which still results in an iterative   scaling procedure and  could be viewed as a generalized IPS.

\subsection{GIS and extensions\label{sub:GIS-and-extensions} }

We derive three surrogate functions all of which are applicable under
the non-negative design setting (\ref{eq:designs}). Recall the objective
function:
\begin{equation}
l(\boldsymbol{\beta})=\sum_{i}[q_{i}\exp(\tilde{\boldsymbol{x}}_{i}^{T}\boldsymbol{\beta})-n_{i}\tilde{\boldsymbol{x}}_{i}^{T}\boldsymbol{\beta}]\triangleq\sum_{i}l_{i}.\label{eq:negative loglikilihood over i}
\end{equation}
Noticing the convexity of $l_{i}$ in $\boldsymbol{\beta}$, we can
apply Jensen's inequality
\begin{align}
l_{i} & =-n_{i}\tilde{\boldsymbol{x}}_{i}^{T}\boldsymbol{\beta}+q_{i}\exp\{\sum_{j}\alpha_{ij}[\frac{x_{ij}}{\alpha_{ij}}(\beta_{j}-\beta_{j}^{-})]+\tilde{\boldsymbol{x}}_{i}^{T}\boldsymbol{\beta}^{-}\}\nonumber \\
 & \leq-n_{i}\tilde{\boldsymbol{x}}_{i}^{T}\boldsymbol{\beta}+\sum_{j}\alpha_{ij}q_{i}\exp[\frac{x_{ij}}{\alpha_{ij}}(\beta_{j}-\beta_{j}^{-})+\tilde{\boldsymbol{x}}_{i}^{T}\boldsymbol{\beta}^{-}]\nonumber \\
 & =-n_{i}\tilde{\boldsymbol{x}}_{i}^{T}\boldsymbol{\beta}+\sum_{j}\alpha_{ij}\mu_{i}^{-}\exp[\frac{x_{ij}}{\alpha_{ij}}(\beta_{j}-\beta_{j}^{-})],\label{eq:Jensen derivation}
\end{align}
where $\alpha_{ij}$ satisfy $\alpha_{ij}\geq0$, $\sum_{j=1}^{p}\alpha_{ij}=1$
for all $i$. We will not allow $\alpha_{ij}$ to be zero when $x_{ij}\neq0$.
If $x_{ij}=0$, the associated term $x_{ij}\beta_{j}$ is $\beta_{j}$-independent and so we set $\alpha_{ij}=0$ for such $x_{ij}$ formally.
This gives
\begin{equation}
l_{i}\leq-n_{i}\tilde{\boldsymbol{x}}_{i}^{T}\boldsymbol{\beta}+\sum_{\{j:x_{ij}\neq0\}}\alpha_{ij}\mu_{i}^{-}\exp[\frac{x_{ij}}{\alpha_{ij}}(\beta_{j}-\beta_{j}^{-})].\label{eq:Jensen's inequality}
\end{equation}

Let's first study binary designs.
With the somewhat naive choice $\alpha_{ij}=1/p$, we immediately
obtain a surrogate satisfying (\ref{eq:MM property}):
\begin{equation}
g_{1}(\boldsymbol{\beta}\mid\boldsymbol{\beta}^{-})=-\boldsymbol{n}^{T}\boldsymbol{X}\boldsymbol{\beta}+\frac{1}{p}\sum_{ij}\mu_{i}^{-}\exp[px_{ij}(\beta_{j}-\beta_{j}^{-})].\label{eq:surr}
\end{equation}
Given any binary design with $x_{ij}=0$ or $1$, which covers the
contingency table setting, the optimal solution of $\textrm{arg}\min_{\boldsymbol{\beta}}g_{1}(\boldsymbol{\beta}\mid\boldsymbol{\beta}^{-})$
is given by:
\[
\beta_{j}=\beta_{j}^{-}+\frac{1}{p}\log[\boldsymbol{x}_{j}^{T}\boldsymbol{n}/(\boldsymbol{x}_{j}^{T}\boldsymbol{\mu}^{-})],\quad j\in[p].
\]
The updates of parameter and mean are obtained as follows
\begin{align}
\boldsymbol{\beta}^{(t+1)} & =\boldsymbol{\beta}^{(t)}+\frac{1}{p}\log[(\boldsymbol{X}^{T}\boldsymbol{n})\oslash(\boldsymbol{X}^{T}\boldsymbol{\mu}^{(t)})]\label{eq:binary_MM_beta}\\
\boldsymbol{\mu}^{(t+1)} & =\boldsymbol{\mu}^{(t)}\circ(\exp\{\frac{1}{p}\boldsymbol{X}\log[(\boldsymbol{X}^{T}\boldsymbol{n})\oslash(\boldsymbol{X}^{T}\boldsymbol{\mu}^{(t)})]\}).\label{eq:binary_MM_mu}
\end{align}

The MM algorithm shares similarities with the IPS defined in Algorithm
\ref{alg:CD alg}, but differs in two ways. IPS updates the components
of $\boldsymbol{\beta}$ sequentially (asynchronously), while (\ref{eq:binary_MM_beta})
updates the entire vector synchronously. Correspondingly, the stepsize
in MM algorithm is smaller than that used in standard IPS.

Now we consider non-negative features.
Recall (\ref{eq:designs}), where we have $x_{i+}\neq0$ for all $i$
from assumption (iii) in Section \ref{subsec:model}. Setting
$\alpha_{ij}=x_{ij}/x_{i+}$ in (\ref{eq:Jensen's inequality}) yields
\begin{equation}
l_{i}\leq-\sum_{j}n_{i}x_{ij}\beta_{j}+\sum_{j}\frac{x_{ij}}{x_{i+}}\mu_{i}^{-}\exp[x_{i+}(\beta_{j}-\beta_{j}^{-})],\label{eq:MM 2 log like non negative}
\end{equation}
which leads to another surrogate function
\[
g_{20}(\boldsymbol{\beta}\mid\boldsymbol{\beta}^{-})=-\boldsymbol{n}^{T}\boldsymbol{X}\boldsymbol{\beta}+\sum_{i,j}\frac{x_{ij}}{x_{i+}}\mu_{i}^{-}\exp[x_{i+}(\beta_{j}-\beta_{j}^{-})].
\]
The optimal $\boldsymbol{\beta}$ satisfies $\sum_{i}x_{ij}\mu_{i}^{-}\exp[x_{i+}(\beta_{j}-\beta_{j}^{-})]=\sum_{i}n_{i}x_{ij}$
but does not enjoy a closed form solution in general.

We majorize the term $\exp[x_{i+}(\beta_{j}-\beta_{j}^{-})]/x_{i+}$
further. For any $a$, $b$ with $0<a\leq b$ and $t\in\mathbb{R}$,
$\exp(at)/a-1/a\leq\exp(bt)/b-1/b$. Applying the inequality with
$a=x_{i+}$ and $b=\max_{i}x_{i,+}=\lVert\boldsymbol{X}\rVert_{\infty}\triangleq R$
($R>0$ from assumption (iii) in Section \ref{subsec:model})
to (\ref{eq:MM 2 log like non negative}) gives
\[
g_{2}(\boldsymbol{\beta}\mid\boldsymbol{\beta}^{-})=-\boldsymbol{n}^{T}\boldsymbol{X}\boldsymbol{\beta}+\sum_{i,j}x_{ij}\mu_{i}^{-}\big\{ \frac{\exp[R(\beta_{j}-\beta_{j}^{-})]}{R}-\frac{1}{R}+\frac{1}{x_{i+}} \big\} .
\]
Now, the optimal solution can be explicitly evaluated: $\beta_{j}=\beta_{j}^{-}+(1/R)\log(\boldsymbol{x}_{j}^{T}\boldsymbol{n}\\/\boldsymbol{x}_{j}^{T}\boldsymbol{\mu}^{-})$,
and the iterates are then given by
\begin{align}
\boldsymbol{\beta}^{(t+1)} & =\boldsymbol{\beta}^{(t)}+\frac{1}{R}\log[(\boldsymbol{X}^{T}\boldsymbol{n})\oslash(\boldsymbol{X}^{T}\boldsymbol{\mu}^{(t)})]\label{eq:non-negative_MM_beta}\\
\boldsymbol{\mu}^{(t+1)} & =\boldsymbol{\mu}^{(t)}\circ(\exp\{\frac{1}{R}\boldsymbol{X}\log[(\boldsymbol{X}^{T}\boldsymbol{n})\oslash(\boldsymbol{X}^{T}\boldsymbol{\mu}^{(t)})]\}).\label{eq:non-negative_MM_mu}
\end{align}
(\ref{eq:non-negative_MM_beta}) is computationally more efficient
than (\ref{eq:binary_MM_beta}) in binary scenarios. (For example,
for a three-way table of size $2\times2\times100$, the main-effects
model has $p=102$ and $R=4$ and so the stepsize in (\ref{eq:non-negative_MM_beta})
is much larger.)
When $R=1$, the update of $\boldsymbol \mu$ corresponds to the GIS  by
\citet{Darroch1972}. (Darroch \& Ratcliff pre-transformed
the original design matrix such that all elements are non-negative
and all row sums are equal to one.)

The MM viewpoint is able to extend GIS, further, to deal with an arbitrary design matrix.
Define the  row support by 
$\mathcal{S}_{j}^{r}\triangleq\{i\in[N]\lvert x_{ij}\neq0\}$, and    $R\triangleq\lVert\boldsymbol{X}\rVert_{\infty}>0$.
Then
$
  g_{3}(\boldsymbol{\beta}\mid\boldsymbol{\beta}^{-})
=  -\boldsymbol{n}^{T}\boldsymbol{X}\boldsymbol{\beta}+\sum_{j\in[p],i\in\mathcal{S}_{j}^{r}}\frac{\mu_{i}^{-}\lvert x_{ij}\rvert}{R}\exp[\frac{x_{ij}}{\lvert x_{ij}\rvert}R(\beta_{j}\\-\beta_{j}^{-})]+\sum_{i}\mu_{i}^{-}(1-\sum_{j}\frac{\lvert x_{ij}\rvert}{R})
$ is a valid surrogate function, and the corresponding
update of $\beta_{j}$ has an explicit expression, which    degenerates to \eqref{eq:non-negative_MM_beta} in the special case of a non-negative design (details omitted).

It is also worth mentioning that MM is capable of deriving
block-wise algorithms that update all blocks in parallel (in contrast to the sequential update of block coordinate descent). For example, consider (\ref{eq:negative loglikilihood over i})
with non-negative designs. Let $\{G_{1},\ldots,G_{k}, \allowbreak\ldots,G_{m}\}$
form a partition of the whole set $[p]$,   $\tilde{\boldsymbol{x}}_{i,k}^{T}$
be the $i$th row vector of the sub-matrix $\boldsymbol{X}_{k}$, $x_{i+}=\langle\boldsymbol{1},\tilde{\boldsymbol{x}}_{i}\rangle$,
$x_{i+,k}=\langle\boldsymbol{1},\tilde{\boldsymbol{x}}_{i,k}\rangle$
and $\alpha_{ik}=x_{i+,k}/x_{i+}$.
Then we can extend   $g_{20}$  to
$
g_{20}'(\boldsymbol{\beta}\mid\boldsymbol{\beta}^{-})=-\boldsymbol{n}^{T}\boldsymbol{X}\boldsymbol{\beta}+\sum_{i,k}\frac{x_{i+,k}}{x_{i+}}\mu_{i}^{-}\exp[\frac{x_{i+}}{x_{i+,k}}\tilde{\boldsymbol{x}}_{i,k}^{T}(\boldsymbol{\beta}_{k}-\boldsymbol{\beta}_{k}^{-})]$. 
A nice feature is that this surrogate is separable in $\boldsymbol{\beta}_{k}$
and so all blocks of coefficients can be simultaneously updated. The convergence of the MM algorithm is typically slower than that
of BCD, but   parallel computing resources should make it   much  more efficient, which  will be investigated in future work. 

\subsection{Reparametrization,    IIS, and quadratic surrogates}
\label{sub:Reparametrization-and-Accelerati}

In this subsection, we assume that $\boldsymbol{X}$ contains a column
corresponding to the intercept, i.e.,
$
\boldsymbol{X}=[\boldsymbol{1}\;\mathring{\boldsymbol{X}}]\in\mathbb{R}^{N\times p}\quad\textrm{and}\quad\boldsymbol{\beta}^{T}=[\beta_{0}\ \mathring{\boldsymbol{\beta}}^{T}]$, 
 where the intercept $\beta_{0}$ is a scalar and and $\mathring{\boldsymbol{\beta}}\in\bar{\mathbb{R}}^{p-1}$
denotes the slope vector. Introducing $\alpha\triangleq\beta_{0}+\log\langle\boldsymbol{q},\exp(\mathring{\boldsymbol{X}}\mathring{\boldsymbol{\beta}})\rangle$,
or $\exp(\alpha)=\langle\boldsymbol{q},\exp(\beta_{0}\boldsymbol{1}+\mathring{\boldsymbol{X}}\mathring{\boldsymbol{\beta}})\rangle=\langle\boldsymbol{1},\boldsymbol{\mu}\rangle$,
$l(\boldsymbol{\beta})$ in (\ref{eq:opt problem MLE beta}) becomes
\begin{eqnarray*}
l(\boldsymbol{\beta}) & = & -\langle\boldsymbol{n},\beta_{0}\boldsymbol{1}+\mathring{\boldsymbol{X}}\mathring{\boldsymbol{\beta}}\rangle+\langle\boldsymbol{q},\exp(\beta_{0}\boldsymbol{1}+\mathring{\boldsymbol{X}}\mathring{\boldsymbol{\beta}})\rangle\\
 & = & -\langle\boldsymbol{n},\mathring{\boldsymbol{X}}\mathring{\boldsymbol{\beta}}\rangle-\langle\boldsymbol{1},\boldsymbol{n}\rangle\beta_{0}+\exp(\alpha)\\
 & = & -\langle\boldsymbol{n},\mathring{\boldsymbol{X}}\mathring{\boldsymbol{\beta}}\rangle-\langle\boldsymbol{1},\boldsymbol{n}\rangle[\alpha-\log\langle\boldsymbol{q},\exp(\mathring{\boldsymbol{X}}\mathring{\boldsymbol{\beta}})\rangle]+\exp(\alpha).
\end{eqnarray*}
The  problem with respect to $\alpha$ can be solved by
$\exp(\alpha)=\langle\boldsymbol{1},\boldsymbol{n}\rangle=\langle\boldsymbol{1},\boldsymbol{\mu}\rangle$,
which means the optimal $\beta_{0}$ satisfies
$
\beta_{0}=\log\langle\boldsymbol{1},\boldsymbol{n}\rangle-\log\langle\boldsymbol{q},\exp(\mathring{\boldsymbol{X}}\mathring{\boldsymbol{\beta}})\rangle
$. 
Therefore, it suffices to study the $\mathring{\boldsymbol{\beta}}$-optimization:
\begin{equation}
\min_{\mathring{\boldsymbol{\beta}}\in\bar{\mathbb{R}}^{p-1}}L(\mathring{\boldsymbol{\beta}})\triangleq-\langle\boldsymbol{n},\mathring{\boldsymbol{X}}\mathring{\boldsymbol{\beta}}\rangle+\langle\boldsymbol{1},\boldsymbol{n}\rangle\log\langle\boldsymbol{q},\exp(\mathring{\boldsymbol{X}}\mathring{\boldsymbol{\beta}})\rangle.\label{eq:opt reparametrized}
\end{equation}

An MM algorithm can be developed for (\ref{eq:opt reparametrized}).
Due to the concavity of the $\log$ function, for any
$\zeta>0$, we have $\log(\zeta x)\leq\zeta x-1$ and this bound is
tight with choice $\zeta(x)=1/x$. Then\begin{equation}
\begin{aligned}L(\mathring{\boldsymbol{\beta}}) & =-\langle\boldsymbol{n},\mathring{\boldsymbol{X}}\mathring{\boldsymbol{\beta}}\rangle+\langle\boldsymbol{1},\boldsymbol{n}\rangle\log[\zeta\langle\boldsymbol{q},\exp(\mathring{\boldsymbol{X}}\mathring{\boldsymbol{\beta}})\rangle]-\langle\boldsymbol{1},\boldsymbol{n}\rangle\log\zeta\\
 & \leq-\langle\boldsymbol{1},\boldsymbol{n}\rangle\log\zeta-\langle\boldsymbol{n},\mathring{\boldsymbol{X}}\mathring{\boldsymbol{\beta}}\rangle+\langle\boldsymbol{1},\boldsymbol{n}\rangle[\zeta\langle\boldsymbol{q},\exp(\mathring{\boldsymbol{X}}\mathring{\boldsymbol{\beta}})\rangle-1].
\end{aligned}
\label{eq:log concavity bound}
\end{equation}
Assuming the non-negative setting (\ref{eq:designs}) and applying
Jensen's inequality (\ref{eq:Jensen's inequality}) with $\alpha_{ij}=\mathring{x}_{ij}/\mathring{x}_{i+}$,
we get
\begin{align}
g_{4}(\mathring{\boldsymbol{\beta}}\mid\mathring{\boldsymbol{\beta}}^{-})=  -\langle\boldsymbol{n},\mathring{\boldsymbol{X}}\mathring{\boldsymbol{\beta}}\rangle-\langle\boldsymbol{1},\boldsymbol{n}\rangle(\log\zeta+1)   +\zeta\langle\boldsymbol{1},\boldsymbol{n}\rangle\sum_{i\in[N],j\in[p-1]}\frac{\mathring{x}_{ij}}{\mathring{x}_{i+}}\mathring{\mu}_{i}^{-}\exp[\mathring{x}_{i+}(\mathring{\beta}_{j}-\mathring{\beta}_{j}^{-})],\label{eq:log-concavity surro}
\end{align}
where $\mathring{\boldsymbol{\mu}}\triangleq\boldsymbol{q}\circ\exp(\mathring{\boldsymbol{X}}\mathring{\boldsymbol{\beta}})$
with its past value denoted by $\mathring{\boldsymbol{\mu}}^{-}=\boldsymbol{q}\circ\exp(\mathring{\boldsymbol{X}}\mathring{\boldsymbol{\beta}}^{-})$.
The only choice to  guarantee that $g_4$ is a surrogate   is $\zeta=1/\langle\boldsymbol{q},\exp(\mathring{\boldsymbol{X}}\mathring{\boldsymbol{\beta}}^{-})\rangle.$
Now, minimizing   (\ref{eq:log-concavity surro}) gives   $\mathring{\beta}_{j}^{(t+1)}$  ($1\le j\le p-1$):
\begin{equation}
\frac{\langle\boldsymbol{1},\boldsymbol{n}\rangle}{\langle\boldsymbol{1},\mathring{\boldsymbol{\mu}}^{(t)}\rangle}\sum_{i}\mathring{x}_{ij}\mathring{\mu}_{i}^{(t)}\exp[\mathring{x}_{i+}(\mathring{\beta}_{j}^{(t+1)}-\mathring{\beta}_{j}^{(t)})]=\sum_{i}n_{i}\mathring{x}_{ij},\quad \label{eq:log-concave update beta}
\end{equation}
and the corresponding proportional scaling on $\mathring{\boldsymbol{\mu}}$ is
given by
\begin{equation}
\mathring{\boldsymbol{\mu}}^{(t+1)}=\mathring{\boldsymbol{\mu}}^{(t)}\circ\exp[\mathring{\boldsymbol{X}}(\mathring{\boldsymbol{\beta}}^{(t+1)}-\mathring{\boldsymbol{\beta}}^{(t)})].\label{eq:log-concave update mu}
\end{equation}
Interestingly, this algorithm can be converted to the celebrated
IIS \citep{Pietra1997} that runs on  the \textit{normalized} observed
and estimated counts, i.e., $\bar{\boldsymbol{n}}\triangleq\boldsymbol{n}/\langle\boldsymbol{1},\boldsymbol{n}\rangle$
and $\bar{\boldsymbol{\mu}}\triangleq\boldsymbol{\mu}/\langle\boldsymbol{1},\boldsymbol{\mu}\rangle$.
\begin{thm}
\label{prop:IIS}The sequence of $\{\mathring{\boldsymbol{\beta}}^{(t)}\}_{t=0}^{\infty}$
generated from the MM algorithm (\ref{eq:log-concave update beta})
and (\ref{eq:log-concave update mu}) coincides with the $\{\mathring{\boldsymbol{\beta}}^{(t)}\}_{t=0}^{\infty}$
generated by IIS:
\begin{align}
\sum_{i}\bar{n}_{i}\mathring{x}_{ij} & =\sum_{i}\mathring{x}_{ij}\bar{\mu}_{i}^{(t)}\exp[\mathring{x}_{i+}(\mathring{\beta}_{j}^{(t+1)}-\mathring{\beta}_{j}^{(t)})],\quad j\in[p-1]\label{eq:IIS_beta}\\
\bar{\boldsymbol{\mu}}^{(t+1)} & =\bar{\boldsymbol{\mu}}^{(t)}\circ\exp[\mathring{\boldsymbol{X}}(\mathring{\boldsymbol{\beta}}^{(t+1)}-\mathring{\boldsymbol{\beta}}^{(t)})]/\langle\bar{\boldsymbol{\mu}}^{(t)},\exp[\mathring{\boldsymbol{X}}(\mathring{\boldsymbol{\beta}}^{(t+1)}-\mathring{\boldsymbol{\beta}}^{(t)})]\rangle.\label{eq:IIS_mu}
\end{align}
\end{thm}

In (\ref{eq:log-concave update beta}) and (\ref{eq:log-concave update mu}), neither the normalization operation nor the intercept update is needed during
the iteration. One can extract the intercept estimate 
at the end.
\\


The reparametrized form (\ref{eq:opt reparametrized}) offers more
options in  surrogate construction. A pleasing fact is that
the term $\langle\boldsymbol{1},\boldsymbol{n}\rangle\log\langle\boldsymbol{q},\exp(\mathring{\boldsymbol{X}}\mathring{\boldsymbol{\beta}})\rangle$
has uniformly bounded curvature. Let's consider  a quadratic
 function $Q$, 
\begin{equation}
Q(\mathring{\boldsymbol{\beta}}\mid\mathring{\boldsymbol{\beta}}^{-})=L(\mathring{\boldsymbol{\beta}}^{-})+\langle\mathring{\boldsymbol{\beta}}-\mathring{\boldsymbol{\beta}}^{-},\nabla_{\mathring{\boldsymbol{\beta}}}L(\mathring{\boldsymbol{\beta}}^{-})\rangle+\frac{1}{2}(\mathring{\boldsymbol{\beta}}-\mathring{\boldsymbol{\beta}}^{-})^{T}\boldsymbol{W}(\mathring{\boldsymbol{\beta}}-\mathring{\boldsymbol{\beta}}^{-}).\label{eq:reparametrize quadratic surrogate}
\end{equation}
By Taylor expansion, $Q$ is a valid surrogate function provided that
\begin{equation}
\langle\boldsymbol{1},\boldsymbol{n}\rangle\mathring{\boldsymbol{X}}^{T}\Big[\frac{\textrm{diag}(\mathring{\boldsymbol{\mu}})}{\langle\boldsymbol{1},\mathring{\boldsymbol{\mu}}\rangle}-\frac{\mathring{\boldsymbol{\mu}}\mathring{\boldsymbol{\mu}}^{T}}{\langle\boldsymbol{1},\mathring{\boldsymbol{\mu}}\rangle^{2}}\Big]\mathring{\boldsymbol{X}}\preceq\boldsymbol{W},\;\forall\mathring{\boldsymbol{\mu}}\succeq\boldsymbol{0}.
\label{eq:Hessian bound}
\end{equation}
 A straightforward choice is
$
\boldsymbol{W}= (\langle\boldsymbol{1},\boldsymbol{n}\rangle\lVert\mathring{\boldsymbol{X}}\rVert_{2}^{2} /2)\,\boldsymbol{I}
$.  It can be refined to
$\boldsymbol{W}=\langle\boldsymbol{1},\boldsymbol{n}\rangle
\mathring{\boldsymbol{X}}^{T}(\boldsymbol{I}-\boldsymbol{1}\boldsymbol{1}^{T}/N)\mathring{\boldsymbol{X}}/2$ \citep{Bohning1988}  noticing the matrix in the brackets of \eqref{eq:Hessian bound} has  an eigenvector $\boldsymbol 1$ associated with eigenvalue $0$.
Based on our experience, the latter 
is better in most large  problems and  will be adopted
unless otherwise specified. In either case, the optimal solution of
$\min_{\mathring{\boldsymbol{\beta}}}Q(\mathring{\boldsymbol{\beta}}\mid\mathring{\boldsymbol{\beta}}^{-})$
is   $\mathring{\boldsymbol{\beta}}^{-}-\boldsymbol{W}^{-1}[-\mathring{\boldsymbol{X}}^{T}\boldsymbol{n}+(\langle\boldsymbol{1},\boldsymbol{n}\rangle/\langle\boldsymbol{1},\mathring{\boldsymbol{\mu}}^{-}\rangle)\mathring{\boldsymbol{X}}^{T}\mathring{\boldsymbol{\mu}}^{-}]$.

Because the last term $(\mathring{\boldsymbol{\beta}}-\mathring{\boldsymbol{\beta}}^{-})^{T}\boldsymbol{W}(\mathring{\boldsymbol{\beta}}-\mathring{\boldsymbol{\beta}}^{-})/2$
in (\ref{eq:reparametrize quadratic surrogate}) can be identified
as a Bregman divergence $\mathbf{D}(\mathring{\boldsymbol{\beta}}\mid\mathring{\boldsymbol{\beta}}^{-})$,
 (\ref{eq:reparametrize quadratic surrogate}) falls into the computational
framework where   Nesterov's second acceleration scheme  applies \citep{nesterov1988}. At each  step, it uses two  auxiliary sequences and adds momentum terms in updating the iterates. It can be shown rigorously that this ingenious approach   leads to improved rate of convergence; see, e.g.,  \cite{Tseng2010}. The resulting  algorithm (Algorithm \ref{alg:QIPS}) is referred to as the \textbf{Q}uadratic-surrogate
\textbf{IPS} (\textbf{Q-IPS}). It is worth mentioning that this momentum-based acceleration does not add much additional cost in each step but  offers significant improvement over IIS and GIS (cf. Section \ref{sec:exp}).
{\begin{algorithm}[t!] \small \caption{\textbf{Q-IPS}. \label{alg:QIPS}}         \textbf{Input} $\boldsymbol{n}$, $\boldsymbol{q}$, $\mathring{\boldsymbol{X}}$\\
\textbf{Initialize }
$\mathring{\boldsymbol{\beta}}^{(0)}\in\mathbb{R}^{p-1}$,\textbf{ $\theta_{0}\leftarrow1$},  $t\leftarrow0$

\begin{algorithmic}[1]
\State
$\mathring{\boldsymbol{\mu}}^{(0)}\leftarrow\boldsymbol{q}\circ\exp(\mathring{\boldsymbol{X}}\mathring{\boldsymbol{\beta}}^{(0)})$, $\mathring{\boldsymbol{\eta}}^{(0)}\leftarrow\mathring{\boldsymbol{\beta}}^{(0)}$
          \While{  not converged }                 \State  $\mathring{\boldsymbol{\alpha}}^{(t)}\leftarrow(1-\theta_{t})\mathring{\boldsymbol{\beta}}^{(t)}+\theta_{t}\mathring{\boldsymbol{\eta}}^{(t)}$                 \State $\mathring{\boldsymbol{\eta}}^{(t+1)}\leftarrow\argmin_{\mathring{\boldsymbol{\beta}}}[\langle\nabla_{\mathring{\boldsymbol{\beta}}}L(\mathring{\boldsymbol{\alpha}}^{(t)}),(\mathring{\boldsymbol{\beta}}-\mathring{\boldsymbol{\alpha}}^{(t)})\rangle+\theta_{t}\mathbf{D}(\mathring{\boldsymbol{\beta}}\mid\mathring{\boldsymbol{\eta}}^{(t)})]$ 
           or
 \Statex $\,\quad$ 
$\mathring{\boldsymbol{\eta}}^{(t+1)}\leftarrow\mathring{\boldsymbol{\eta}}^{(t)}-\theta_{t}^{-1}\boldsymbol{W}^{-1}\{-\mathring{\boldsymbol{X}}^{T}\boldsymbol{n}+\frac{\langle\boldsymbol{1},\boldsymbol{n}\rangle}{\langle\boldsymbol{q}, \exp(\mathring{\boldsymbol{X}}\mathring{\boldsymbol{\alpha}}^{(t)})\rangle}\mathring{\boldsymbol{X}}^{T}[\boldsymbol{q}\circ\exp(\mathring{\boldsymbol{X}}\mathring{\boldsymbol{\alpha}}^{(t)})]\} $
 \State $\mathring{\boldsymbol{\beta}}^{(t+1)}\leftarrow(1-\theta_{t})\mathring{\boldsymbol{\beta}}^{(t)}+\theta_{t}\mathring{\boldsymbol{\eta}}^{(t+1)}$ \State $\mathring{\boldsymbol{\mu}}^{(t+1)}\leftarrow\mathring{\boldsymbol{\mu}}^{(t)}\circ\exp[\mathring{\boldsymbol{X}}(\mathring{\boldsymbol{\beta}}^{(t+1)}-\mathring{\boldsymbol{\beta}}^{(t)})]$ \State $\theta_{t+1}\leftarrow(\sqrt{\theta_{t}^{4}+4\theta_{t}^{2}}-\theta_{t}^{2})/2$ \State $t\leftarrow t+1$ \EndWhile \State$\beta_{0}^{(t)}\leftarrow\log\langle\boldsymbol{1},\boldsymbol{n}\rangle-\log\langle\boldsymbol{q},\exp(\mathring{\boldsymbol{X}}\mathring{\boldsymbol{\beta}}^{(t)})\rangle$ \State $\boldsymbol{\mu}^{(t)}\leftarrow\exp(\beta_{0}^{(t)})\mathring{\boldsymbol{\mu}}^{(t)}$ \\  \Return $\hat{\boldsymbol{\mu}}=\boldsymbol{\mu}^{(t)}$, $\hat{\boldsymbol{\beta}}=[\beta_{0}^{(t)}\ \mathring{\boldsymbol{\beta}}^{(t)T}]^{T}$         \end{algorithmic} \end{algorithm}}


\section{Regularized Estimation}
\label{sec:Shrinkage-estimation}

\noindent Modern statistical applications often involve a large number
of variables, where regularization is necessary to achieve
estimation accuracy or model parsimony. For example, one can append
an $\ell_{2}$-type penalty to the negative log-likelihood to handle
collinearity:
\noindent
\begin{equation}
\min_{}f_{2}(\boldsymbol{\beta})\triangleq-\langle\boldsymbol{n},\boldsymbol{X}\boldsymbol{\beta}\rangle+\langle\boldsymbol{q},\exp(\boldsymbol{X}\boldsymbol{\beta})\rangle+\frac{\lambda}{2}\sum_{j=2}^p \beta_j^2, \label{eq:opt problem MLE beta l2}
\end{equation}
where $\beta_1$ corresponds to the intercept column and $\lambda\ge 0$ is a regularization parameter.
$\lambda$   can be tuned
by  AIC \citep{akaike1974} but   even fixing it
at a small value (say $1\textrm{e-}5$) often shows improved accuracy.   We recommend including such a mild $\ell_{2}$-penalty in
practical applications, especially those with zero counts.
 Many of the previously
developed algorithms   can be easily modified
to adapt to \eqref{eq:opt problem MLE beta l2} and the details are
not discussed.

Another popular way of regularization is to enforce sparsity, 
which can help practitioners select a small set of relevant features.
 The coordinate descent characterization of IPS enables us to  develop its sparse variants on contingency tables.
Assume a binary design $\boldsymbol{X}$ (cf. (\ref{eq:designs})) and
 consider the following problem subject to an $\ell_1$- or $\ell_0$-penalty
\begin{equation}
\min_{\boldsymbol{\beta}} f_{1}(\boldsymbol{\beta}) \,(\mbox{or } f_{0}(\boldsymbol{\beta})) \triangleq-\langle\boldsymbol{n},\boldsymbol{X}\boldsymbol{\beta}\rangle+\langle\boldsymbol{q},\exp(\boldsymbol{X}\boldsymbol{\beta})\rangle+\sum_{j=1}^p \lambda_j | \beta_j|  \,(\mbox{or }\sum_{j=1}^p\lambda_j 1_{\beta_j\ne 0}),\label{eq:opt problem MLE beta l1}
\end{equation}
where typically $\lambda_1 = 0$ and $\lambda_j=\lambda$ for $2\le j \le p$.
The following theorem can be used to   derive the coordinate-wise
update for (\ref{eq:opt problem MLE beta l1}).
\begin{thm}
\label{lem:l1 univariate}Let $\boldsymbol{n},\boldsymbol{x}, \boldsymbol q\in\mathbb{\mathbb{R}}^{N}$,  $\beta\in\mathbb{\mathbb{R}}$, $\lambda\ge 0$. Then the solution to the optimization
problem $\min_{\beta\in\mathbb{\mathbb{R}}}-\langle\boldsymbol{n},\allowbreak\boldsymbol{x}\rangle \beta+\langle\boldsymbol{q},\exp(\beta\boldsymbol{x})\rangle+\lambda\lvert\beta\rvert$
is given by $\hat{\beta}=\log\{[\langle\boldsymbol{x},\boldsymbol{n}\rangle-\lambda\sgn(\langle\boldsymbol{x},\boldsymbol{n}-\boldsymbol{q}\rangle)]/\langle\boldsymbol{x},\boldsymbol{q}\rangle\}$
if $\lvert\langle\boldsymbol{x},\boldsymbol{n}-\boldsymbol{q}\rangle\rvert\geq\lambda$
and $\hat{\beta}=0$ otherwise. Also, a global solution to  $\min_{\beta\in\mathbb{\mathbb{R}}}-\langle\boldsymbol{n},\allowbreak\boldsymbol{x}\rangle \beta+\langle\boldsymbol{q},\exp(\beta\boldsymbol{x})\rangle+\lambda1_{\beta \ne 0}$ is $\hat \beta = \log\{\langle\boldsymbol{x},\boldsymbol{n}\rangle/\langle\boldsymbol{x},\boldsymbol{q}\rangle\}$
if $\DKL(\langle\boldsymbol{x},\boldsymbol{n}\rangle \| \langle\boldsymbol{x},\boldsymbol{q}\rangle)\ge \lambda $ and $0$ otherwise.
\end{thm}
We use the $\ell_1$-penalized problem   as an example to show  how to modify  Algorithm  \ref{alg:CD alg}  to get a sparsity-pursuing IPS  (a similar algorithm can be developed in the $\ell_0$ case).
Let $\boldsymbol{\mu}^{(t,j-1)}=\boldsymbol{q}\,\circ\,\exp(\boldsymbol{X}[\beta_{1}^{(t+1)},\ldots,\beta_{j-1}^{(t+1)},\beta_{j}^{(t)}\ldots,\beta_{p}^{(t)}]^{T})$.
From  Theorem \ref{lem:l1 univariate},
for any   $j:1\leq j\leq p$$
, \beta_{j}^{(t+1)}=\arg  \allowbreak \min_{\beta_{j}}\allowbreak f_{1} (\beta_{1}^{(t+1)}, \allowbreak \ldots ,\allowbreak  \beta_{j-1}^{(t+1)},\allowbreak \beta_{j},\beta_{j+1}^{(t)}\ldots,\beta_{p}^{(t)})$, or $
\beta_{j}^{(t)}+\log[\frac{\langle\boldsymbol{x}_{j},\boldsymbol{n}\rangle-\lambda_{j}\sgn(\delta_{t,j})}{\langle\boldsymbol{x}_{j},\boldsymbol{\mu}^{(t,j-1)}\rangle}] $ if  $\lvert\delta_{t,j}\rvert\geq\lambda_{j}$
and $0$ otherwise, 
 $\delta_{t,j}\triangleq\langle\boldsymbol{x}_{j},\boldsymbol{n}-\boldsymbol{\mu}^{(t,j-1)}\circ\exp(-\beta_{j}^{(t)}\boldsymbol{x}_{j})\rangle=\langle\boldsymbol{x}_{j},\boldsymbol{n}-\boldsymbol{\mu}^{(t,j-1)}\exp(-\beta_{j}^{(t)})\rangle$.
When the intercept is  not subject to any penalty, it  can be updated by $\beta_{1}^{(t+1)}=\beta_{1}^{(t)}+\log(\langle\boldsymbol{x}_{1},\boldsymbol{n}\rangle/\langle\boldsymbol{x}_{1},\boldsymbol{\mu}^{(t,0)}\rangle)$. 
So the iterative scaling on the mean vector  is
$\boldsymbol{\mu}^{(t,j)}=\boldsymbol{\mu}^{(t,j-1)}\circ\exp[\boldsymbol{x}_{j}(\beta_{j}^{(t+1)}-\beta_{j}^{(t)})]$, or $\boldsymbol{\mu}^{(t,j)}=\boldsymbol{\mu}^{(t,j-1)}\circ\exp\{\boldsymbol{x}_{j}\\\log[(\boldsymbol{x}_{j}^{T}\boldsymbol{n}-\lambda_{j}\sgn(\langle\boldsymbol{x}_{j},\boldsymbol{n}-\boldsymbol{\nu}^{(t,j-1)}\rangle))/(\boldsymbol{x}_{j}^{T}\boldsymbol{\mu}^{(t,j-1)})]\}$ if $|\langle\boldsymbol{x}_{j},\boldsymbol{n}-\boldsymbol{\nu}^{(t,j-1)}\rangle|\ge \lambda_j$  and $\boldsymbol{\nu}^{(t,j-1)}$ otherwise, where ${\boldsymbol  \nu}^{(t, j-1)} =\boldsymbol{q}\,\circ\,\exp(\boldsymbol{x}_1 \beta_{1}^{(t+1)}+\cdots+\boldsymbol{x}_{j-1}\beta_{j-1}^{(t+1)}+ \boldsymbol{x}_{j+1}\beta_{j+1}^{(t)}+\cdots+ \boldsymbol{x}_p\beta_{p}^{(t)})$. This  modified Algorithm  \ref{alg:CD alg} is termed \textbf{$\boldsymbol{\ell_{1}}$-IPS}.

In contrast to the  greedy procedures for maximum entropy with concurrent feature selection in the literature, our
 optimization-based algorithm   is stable in the sense that there is an associated objective function    \eqref{eq:opt problem MLE beta l1}  that is guaranteed to be non-increasing, and no quadratic approximation or line
search is required. 

\section{Experiments }\label{sec:exp}


This section tests various algorithms on both contingency tables and log-linear models with non-binary features.
The observed vector   $\boldsymbol{n}=[n_{i}]$  has independent entries satisfying $n_{i}\sim\textrm{Poi}(\mu_{i}^*)$ ($1\le i \le N$), where  the mean vector
$\boldsymbol{\mu}^{*}= \exp(\boldsymbol{X}\boldsymbol{\beta}^{*})$ and   the first column of the design matrix $ \boldsymbol{X}$ is assumed to be $\boldsymbol{1}$. The details  of  how to generate the true coefficient vector     $\boldsymbol{\beta}^{*}$ will vary in different setttins.   Given each setting,
we simulate $20$   data sets and report averaged results.  The measures to characterize the  computational  \textit{optimization  error}
and  statistical  \textit{estimation  error} of the $t$-th iterate
are    the relative gradient $\lVert\boldsymbol{g}_{t}\rVert_{\infty}/\lVert\boldsymbol{g}_{\textrm{0}}\rVert_{\infty}$,
and   the relative estimation error $\lVert\boldsymbol{\beta}^{(t)}-\boldsymbol{\beta}^{*}\rVert_{2}^{2}/\lVert\boldsymbol{\beta}^{*}\rVert_{2}^{2}$,
where $\boldsymbol{g}_{t}$ ($t\geq0$) is the gradient of the original
objective function (\ref{eq:opt problem MLE beta}) evaluated at $\boldsymbol{\beta}^{(t)}$.
   The termination
criterion is met if
$\lVert\boldsymbol{g}_{t}\rVert_{\infty} / \lVert\boldsymbol{g}_{0}\rVert_{\infty}\leq\epsilon_{\textrm{tol}} $ 
  or the running time  is greater than $t_{\max}$. By
default, $\epsilon_{\textrm{tol}}=1\textrm{e-}4$ and $t_{\max}=600$. Typically, we will plot   computational and statistical errors
 (on the log scale for better  visibility)  against  computational time, rather than the number of iterations,  since different algorithms may have very different per-iteration complexity.

The algorithms for comparison include IPS (cf. Alg.  \ref{alg:CD alg}),   A-IPS (cf. Alg.  \ref{alg:AIPS}),    GIS (cf. \eqref{eq:non-negative_MM_beta} \& \eqref{eq:non-negative_MM_mu}),  IIS (cf.  \eqref{eq:IIS_beta} \&  \eqref{eq:IIS_mu}),   Newton \citep{LBFGS2012},  Q-IPS (cf. Alg.  \ref{alg:QIPS}), and  B-IPS  on the reparametrized problem (\ref{eq:opt reparametrized}).
The initial value $\boldsymbol{\beta}^{(0)}$   is    $\boldsymbol{0}$ in all experiments.
The block sizes in B-IPS are set to be $g_{k}=200$ unless otherwise
specified.  An efficient implementation of Newton's method
with bracketing line-search   can be found in   \cite{LBFGS2012}. 
The simulations   are conducted
on   a PC  with   2.9GHz CPU, 16GB memory, and 64-bit Windows 10.
In the  following, we    first run  IPS algorithms on contingency tables to examine the power of acceleration brought by different randomization strategies, then compare IIS, GIS, Newton-type algorithms and our proposed generalized IPS on   problems with non-binary features, as well as  studying their   scalability on large-scale data. At the end,   a real marketing campaign dataset is analyzed with the sparsity-pursuing   IPS.
%

\subsection{Power of randomization }

\begin{figure}[H]
\begin{centering}
\subfloat[\scriptsize Setting (i)]
{\begin{centering}
\includegraphics[width=0.4\columnwidth]{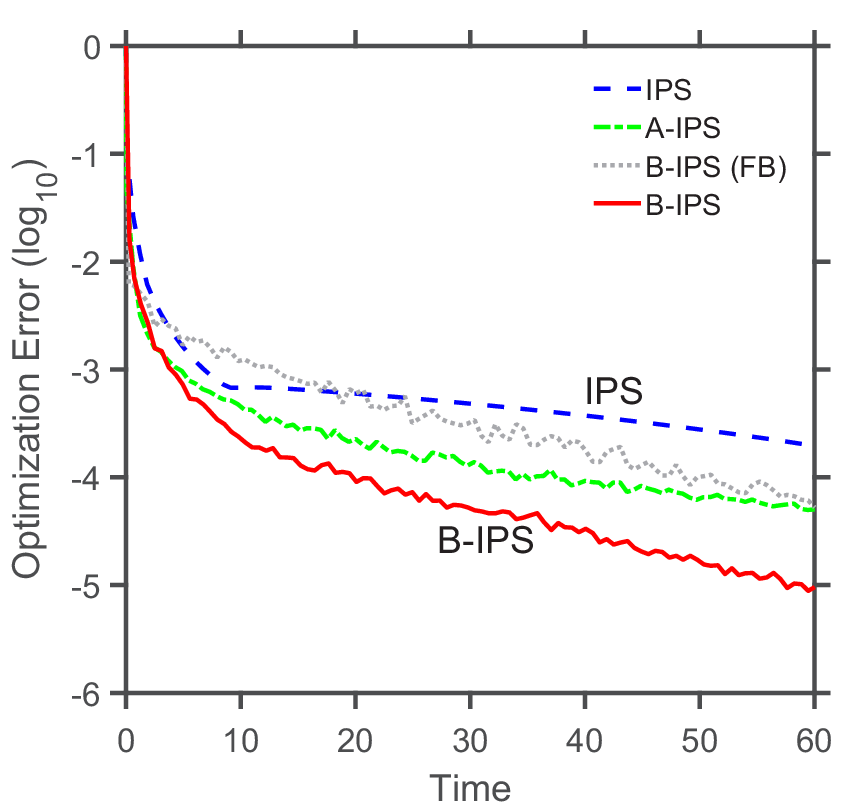}\includegraphics[width=0.4\columnwidth]{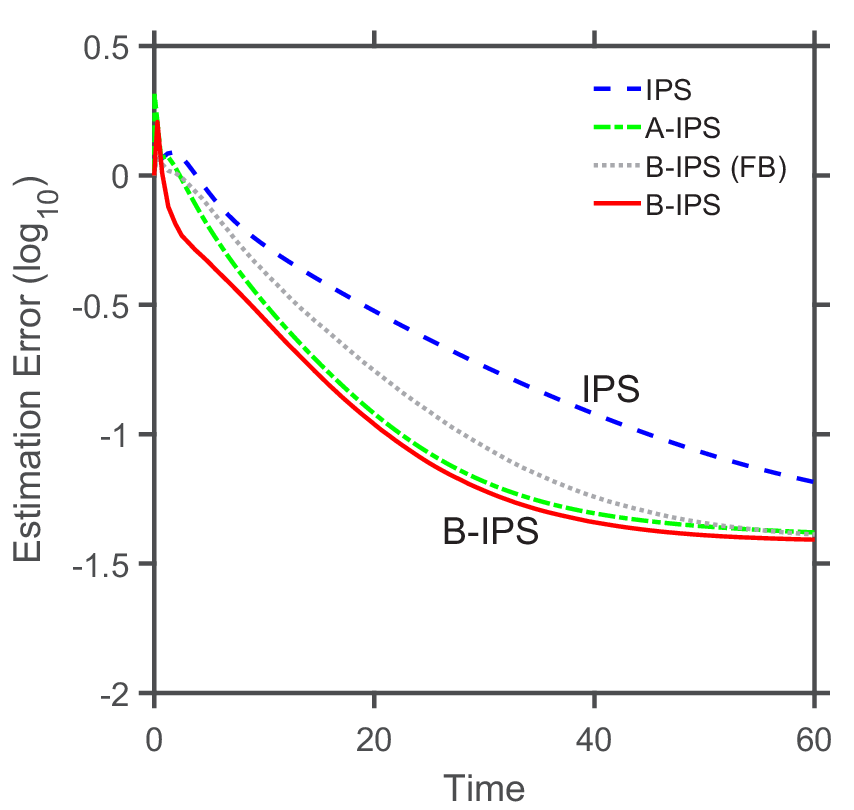}
\par\end{centering}

}
\par\end{centering}

\centering{}
\subfloat[\scriptsize Setting (ii)]
{\begin{centering}
\includegraphics[width=0.4\columnwidth]{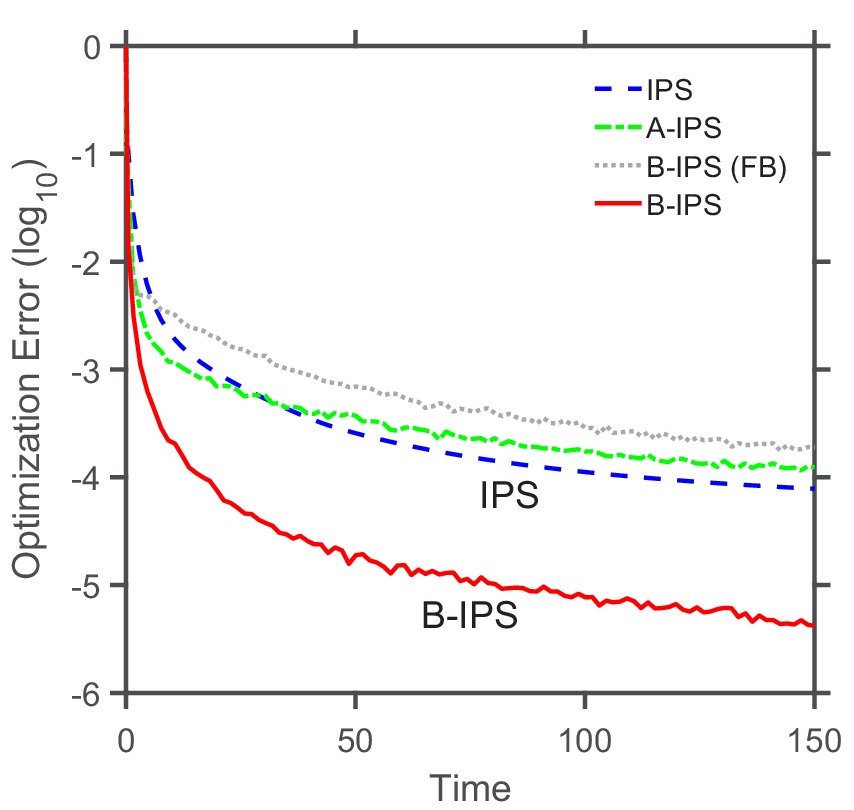}\includegraphics[width=0.4\columnwidth]{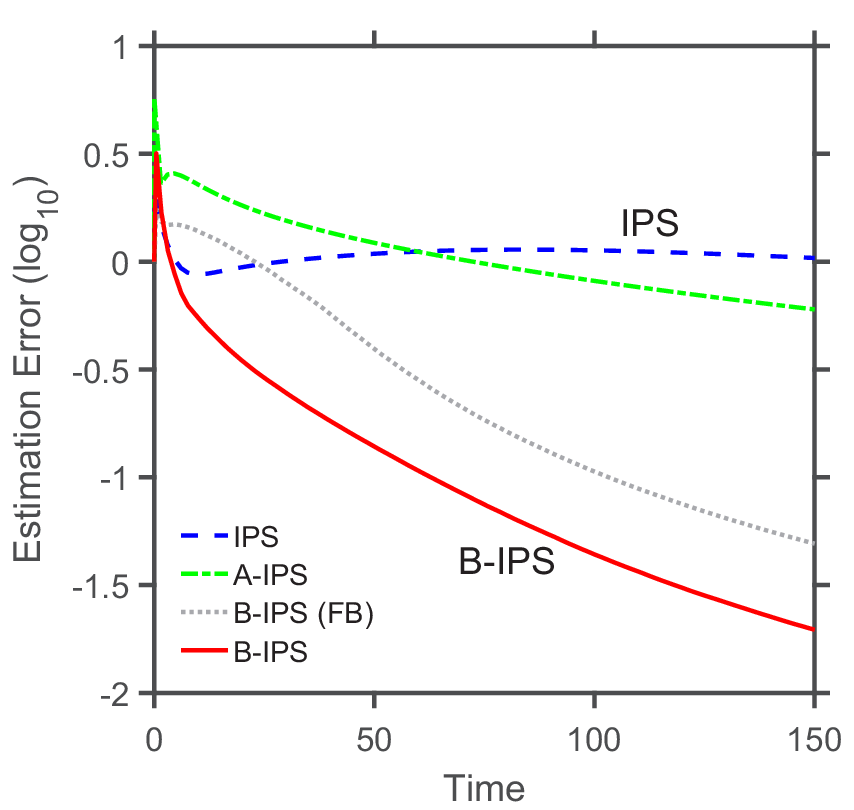}
\par\end{centering}
}
\caption{\footnotesize Performance comparison between IPS, A-IPS,  B-IPS (FB) (fixed blocking, resampling  without replacement) and B-IPS (random blocking followed by cyclic update) on   $10\times10\times10\times10$ tables under a homogeneous association model  ($p=523$). The optimization and estimation errors  are shown on the \textbf{log} scale.  \label{fig:Comparisons-of-progress-table-small}}

\end{figure}

The first experiment   compares different random schemes  on a $10\times10\times10\times10$ table with all  two-way interactions.
This  homogeneous association model  \citep{Agresti2012} has 523 predictors, including $36$ main effects
and $486$ two-way associations in addition to the intercept.
The coefficient vector is generated in two ways. In the first setting, the  intercept is $2$,  the last 10 components are  sampled independently from $0.5\mathcal{N}(1,1)+0.5\mathcal{N}(3,1)$,   and  the remaining are  $0$;
in the second setting, we set 30 randomly chosen components of $\beta_j^*$ to nonzero, sampled from the previous mixture distribution, and set the rest zero.
The results averaged over $20$ runs are shown in Figure \ref{fig:Comparisons-of-progress-table-small}.

According to the figure,  A-IPS is much faster than IPS in the first setting, and
is comparable to it in the second setting, but in both scenarios, A-IPS delivers
far more statistically accurate estimates. (The optimization and estimation errors  are shown on the {log} scale.) Perhaps surprisingly, fixed blocking with random block selection does not show the full power of randomization.
Our  random-blocking-based B-IPS is clearly the winner both  in computational efficiency
and in statistical estimation.

\begin{figure}[ht]
\centering{}\includegraphics[width=0.4\columnwidth]{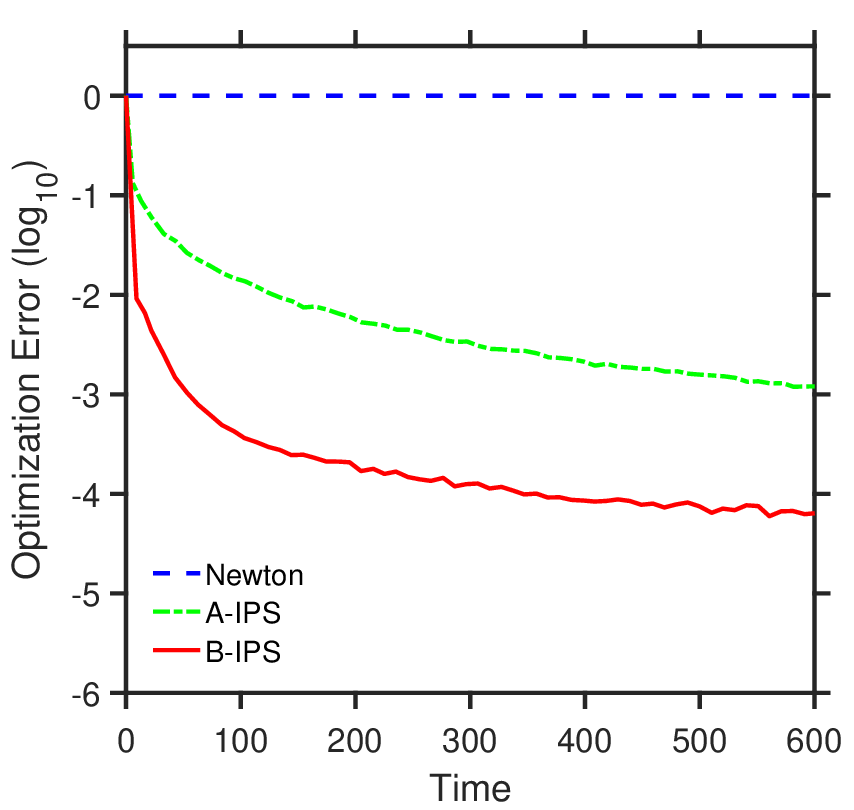}\includegraphics[width=0.4\columnwidth]{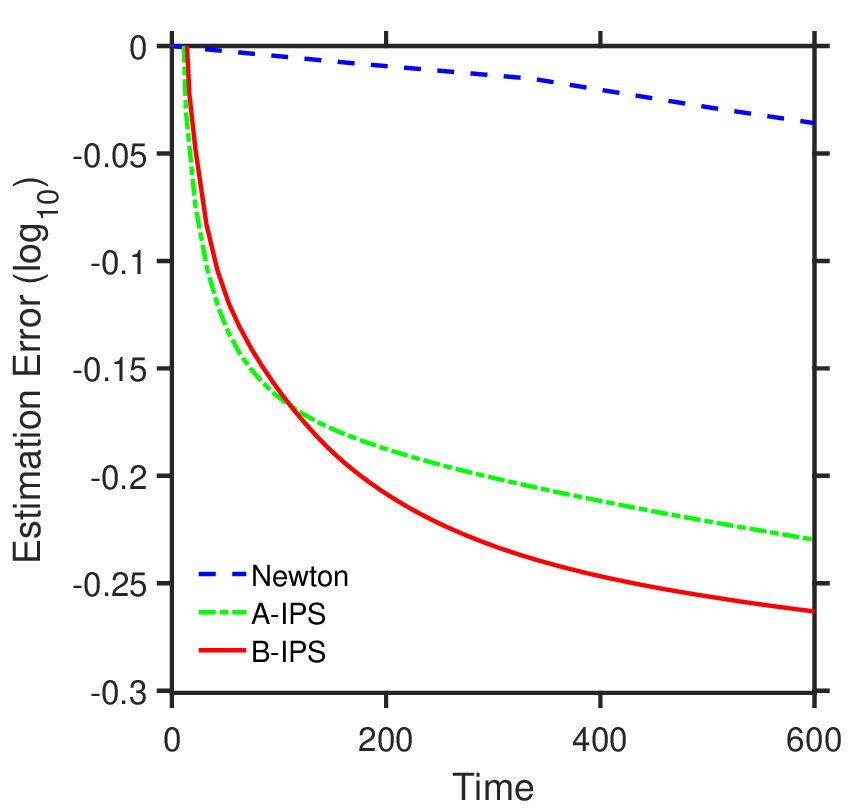}\caption{\footnotesize Optimization error and estimation error (on the scale of \textbf{log}$_{10}$) of Newton, A-IPS, and B-IPS
  for a three-way homogeneous association model on   $10\times10\times10\times10\times10$ tables ($p=8,\negthinspace146$, $N=100,\negthinspace000$).
\label{fig:Comparisons-of-progress-table-large}}
\end{figure}

Next,  we compare  A-IPS and B-IPS
with  Newton's method on  a table of size $10\times10\times10\times10\times10$, where
 a three-way homogeneous association model including \textit{all}
interactions up to third order is assumed (and thus $p=8,\negthinspace146$). The coefficients are generated such that  $\beta_{0}^{*}=5$, the last $2,\negthinspace000$  are independently sampled from
$\mathcal{N}(1,1)$ and the rest  zero.
Seen from  Figure  \ref{fig:Comparisons-of-progress-table-large}, Newton's
algorithm could not deliver a useful estimate within the time limit. It was quite memory-demanding in the experiment. In
comparison, A-IPS and B-IPS offer good scalability and statistical
accuracy.

\subsection{Generalized IPS algorithms}
\label{sub:Log-linear-models:-generalizatio}

This subsection  goes beyond the binary feature setting and   we use generalized IPS to handle nonnegative features as well as  those having both positive and negative values. By default, the prototype design $\mathring{\boldsymbol{X}}$
(the submatrix of $\boldsymbol X$ excluding the intercept) is generated with each row sampled independently
from $\mathcal{N}(\boldsymbol{0},[\rho^{\lvert j-k\rvert}])$, where
$\rho=0.8$ and $1\leq j,k\leq p-1$. Some shifting and scaling operations
will be performed to avoid
overflow issues or  to make the design  non-negative.

\begin{figure}[H]
\begin{centering}
\subfloat[\scriptsize Optimization \& statistical errors against time]{\begin{centering}
\includegraphics[width=0.4\columnwidth]{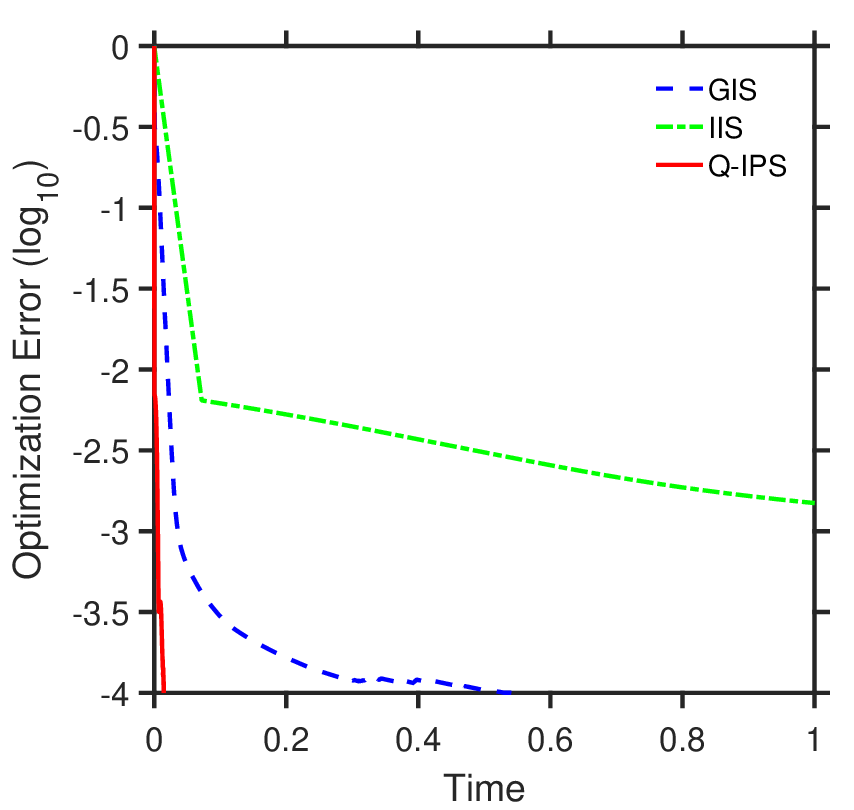}
\includegraphics[width=0.4\columnwidth]{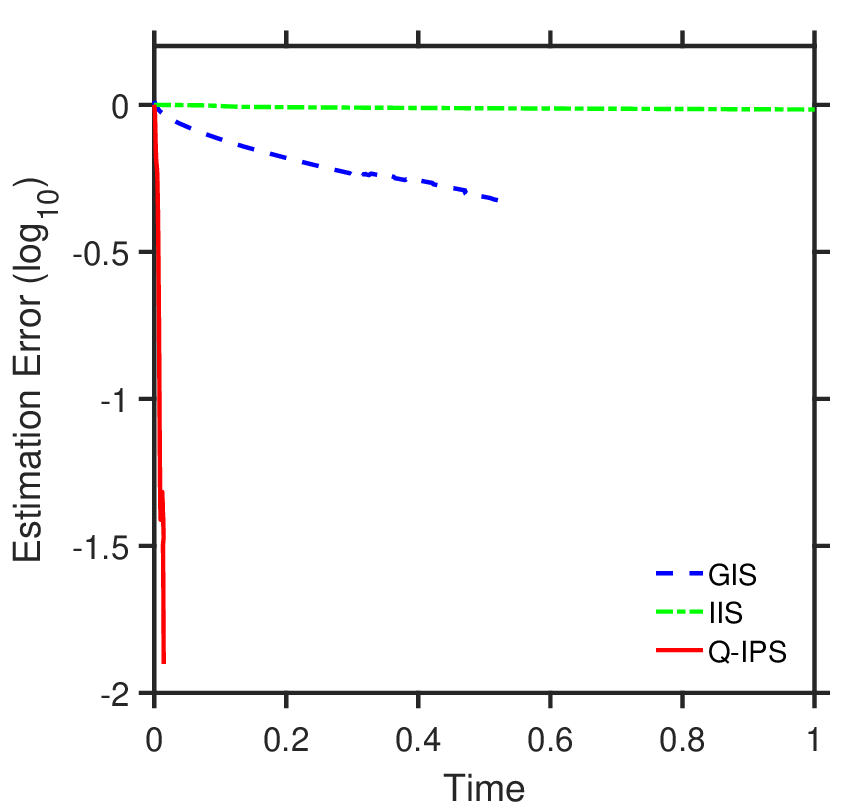}
\par\end{centering}

}
\par\end{centering}

\begin{centering}
\subfloat[\scriptsize Optimization \& statistical errors against iteration]{\begin{centering}
\includegraphics[width=0.4\columnwidth]{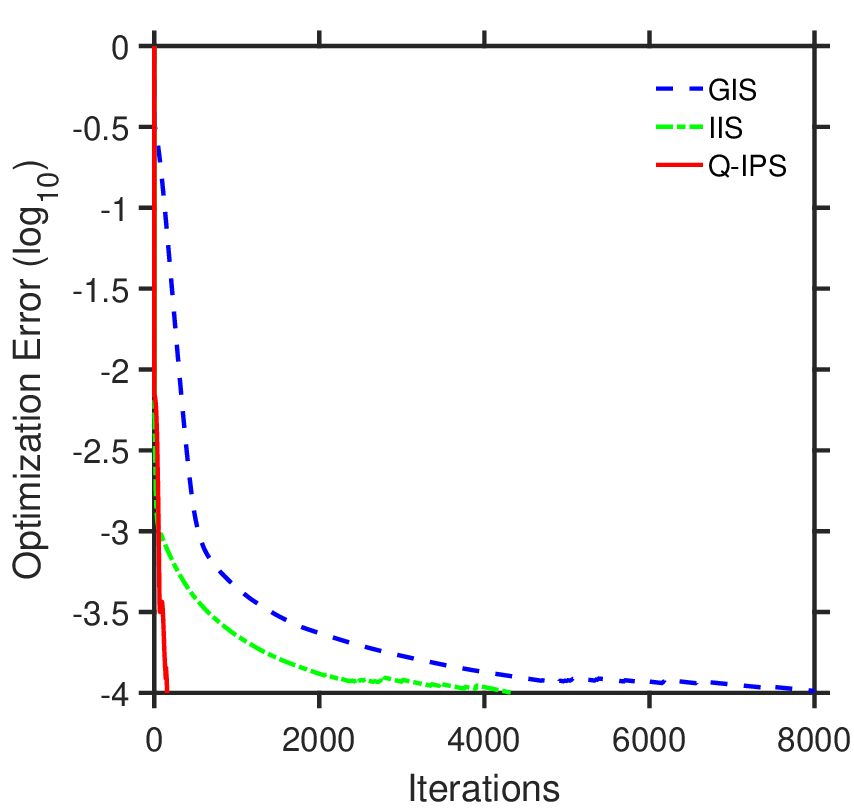}
\includegraphics[width=0.4\columnwidth]{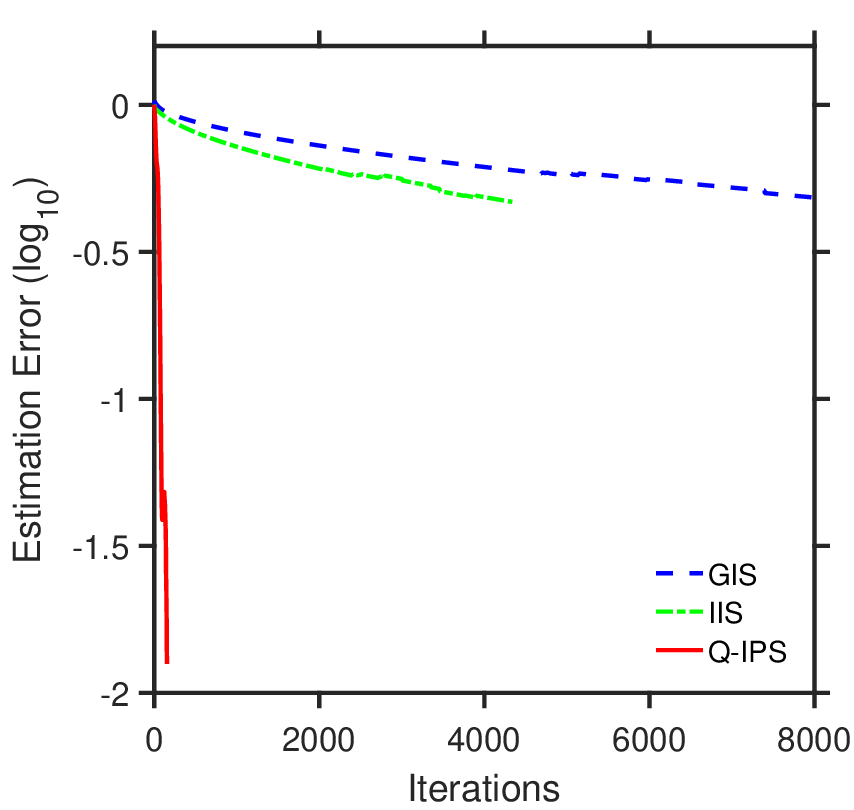}
\par\end{centering}

}
\par\end{centering}

\caption{\footnotesize Performance comparison between GIS, IIS and Q-IPS  ($N=1,\negthinspace000$, $p=100$). The top two plots show how optimization and estimator  errors change over time, and the bottom two  are against the iteration number.  \label{fig:Comparisons-of-progress-nn}}

\end{figure}

The first experiment is to find the winning algorithm among   GIS,  IIS, and Q-IPS. The celebrated GIS and IIS only run on non-negative features, which can be produced by $\mathring{\boldsymbol{X}}\leftarrow\mathring{\boldsymbol{X}}-(\textrm{min}_{i,j}\mathring{x}_{ij})\boldsymbol{1}\boldsymbol{1}^{T}$.
We scale it down further,  $\mathring{\boldsymbol{X}}\leftarrow\mathring{\boldsymbol{X}}/(50\lVert\mathring{\boldsymbol{X}}\rVert_{\max})$,  for better numerical stability.
An artifact  is that the obtained matrix has all row sums approximately equal, which appears too ideal in the real-world. So we
multiply each row by a random factor $1+\lvert z_{i}\rvert$ with
$z_{i}\stackrel{\text{i.i.d}}{\sim}\mathcal{N}(0,1)$. The true coefficient
vector is generated by  $\beta_{0}^{*}=1$ and $\beta_{j}^{*}\stackrel{\text{i.i.d}}{\sim}0.5\mathcal{N}(10,1)+0.5\mathcal{N}(-10,1)$
for $1\leq j\leq p-1$.
The results  shown in Figure \ref{fig:Comparisons-of-progress-nn} are for  $N=1,\negthinspace000$ and
$p=100$, where we plot the algorithm progress (in terms of optimization error and estimation error) as the  computational time
 or iteration number increases.
(To better differentiate GIS and Q-IPS, we only show part of the plots.)
Although  IIS needs fewer  iterations, GIS is found to be more efficient than IIS. This is due to the  cost of  solving $p-1$ nonlinear equations at each iteration of IIS. 
We also see that   Q-IPS\textsuperscript{} outperforms GIS and IIS.


\begin{figure}[h]
\begin{centering}
\subfloat[\scriptsize $p=2,000$ \label{fig:Comparisons-of-progress-nn-scalability-a}]{\begin{centering}
\includegraphics[width=0.4\columnwidth]{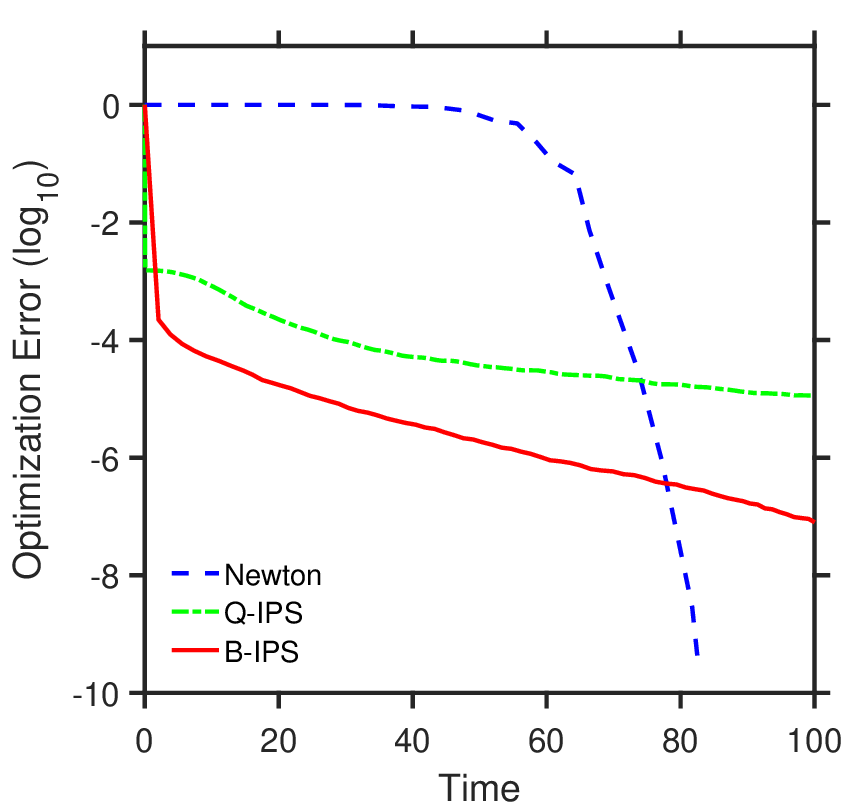}
\includegraphics[width=0.4\columnwidth]{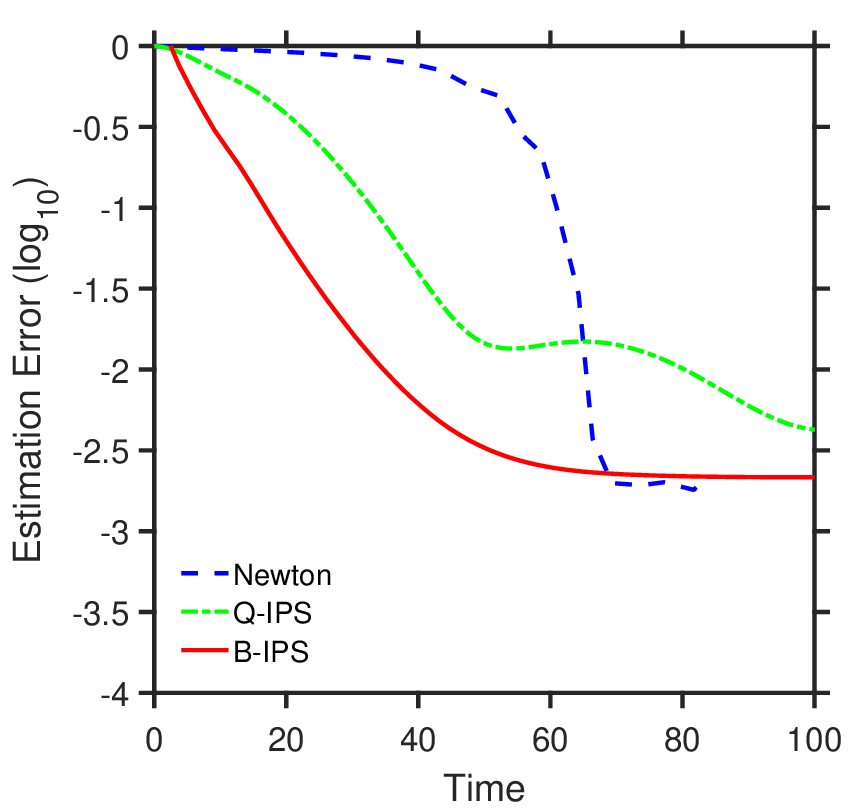}
\par\end{centering}

}
\par\end{centering}

\begin{centering}
\subfloat[\scriptsize $p=4,000$ \label{fig:Comparisons-of-progress-nn-scalability-b}]{\begin{centering}

\includegraphics[width=0.4\columnwidth]{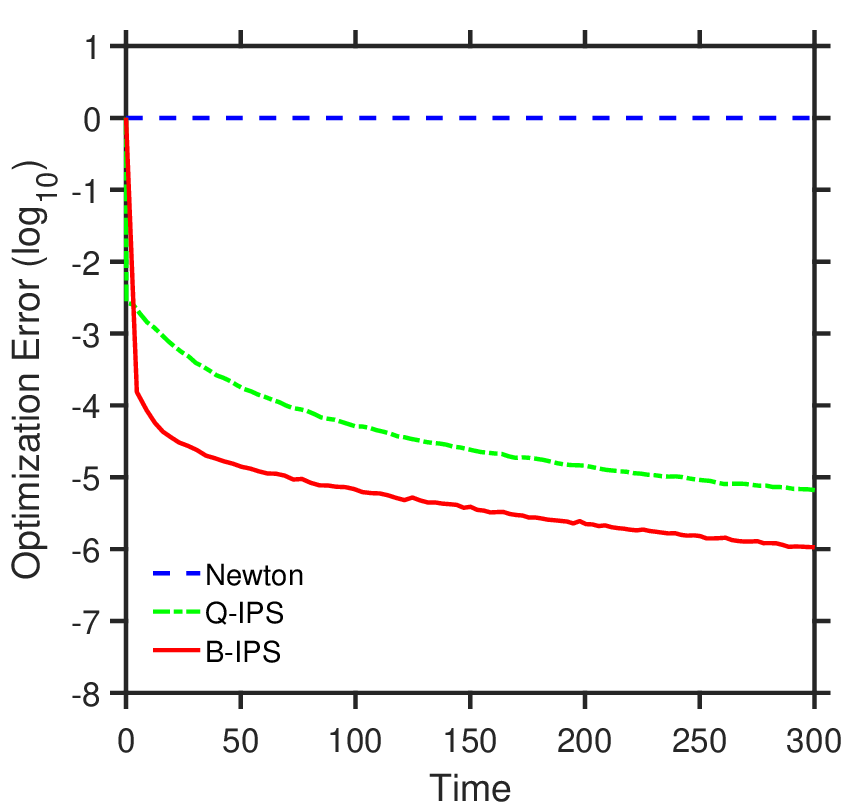}
\includegraphics[width=0.4\columnwidth]{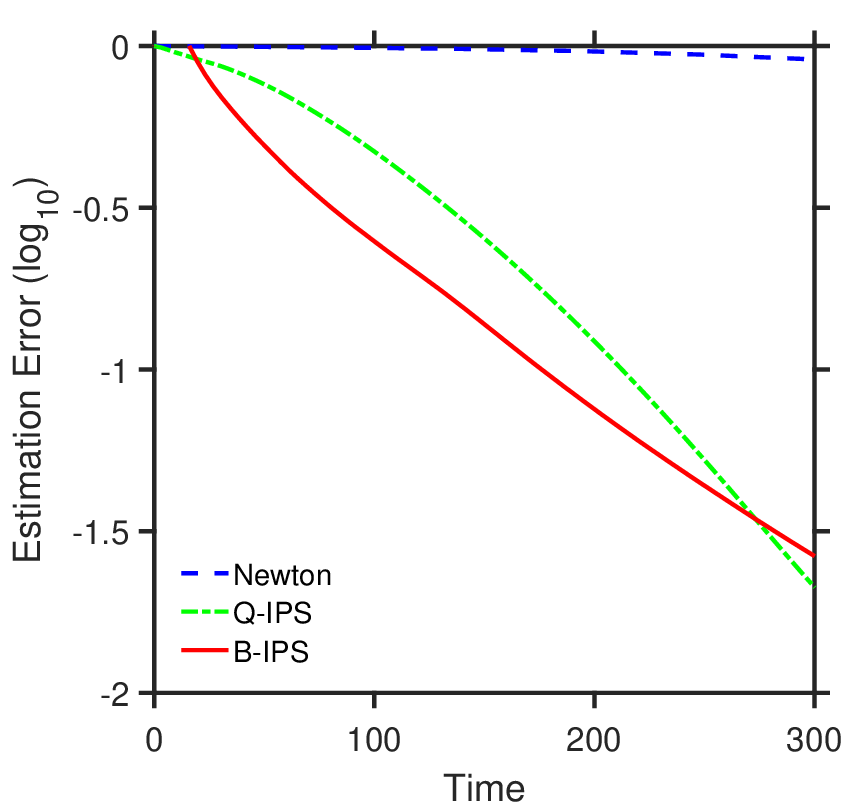}
\par\end{centering}

}
\par\end{centering}
\caption{\footnotesize Performances comparison between Q-IPS, B-IPS and Newton on large non-negative
designs with $N=20,\negthinspace000$, $p=2,\negthinspace000$ (top), $4,\negthinspace000$ (bottom). \label{fig:Comparisons-of-progress-nn-scalability}}

\end{figure}

Next, we conduct a larger experiment with   $p=2,\negthinspace000, 4,\negthinspace000$ and $N=20,\negthinspace000$ , to compare Q-IPS and B-IPS with
Newton.    In generating the simulation data, we  scale down the
prototype design by $\mathring{\boldsymbol{X}}\leftarrow\mathring{\boldsymbol{X}}/(20\lVert\mathring{\boldsymbol{X}}\rVert_{\max})$,
and set $\beta_{j}^{*}\stackrel{\text{i.i.d}}{\sim}0.5\mathcal{N}(10,1)+0.5\mathcal{N}(-10,1)$
for $1\leq j\leq p-1$ and $\beta_{0}^{*}=10$. 
Figure \ref{fig:Comparisons-of-progress-nn-scalability-a})
demonstrates    two typical stages of  Newton's method: the \textit{damped Newton}
phase and the \textit{quadratically convergent} phase (see \cite{Boyd2004}). Though
converging super fast in the second stage, in   \ref{fig:Comparisons-of-progress-nn-scalability-b}), this algorithm took too long
to complete the first stage,  thereby perhaps
less useful in big data applications. In contrast, B-IPS and Q-IPS
are able to deliver accurate estimates within the  time limit, and
B-IPS seems to have better scalability than Q-IPS.

Finally, we turn  to problems with   features not restricted to be nonnegative and compare the performance of Newton, B-IPS and L-BFGS. We set $\beta_{0}^{*}=10$ and $\beta_{j}^{*}\stackrel{\text{i.i.d}}{\sim}0.5\mathcal{N}(10,1)+0.5\mathcal{N}(-10,1)$
for $1\leq j\leq p-1$ and scale the prototype design matrix
by $\mathring{\boldsymbol{X}}\leftarrow\mathring{\boldsymbol{X}}/(100\lVert\mathring{\boldsymbol{X}}\rVert_{\max})$.
Fixing $N=50,\negthinspace000$, we vary $p$ from $1,\negthinspace000$
to $12,\negthinspace000$.
When $p=10,\negthinspace000$ or $12,\negthinspace000$,   not
all methods converge fast, and we set a time limit $t_{\textrm{max}}=600$.
The relative gradient (relGrad) and the estimation error (estErr)
are scaled by $1\textrm{e+}7$ and $1\textrm{e+}4$, respectively, when reported  in Table \ref{tab:general}.

\begin{table}[h]
\footnotesize
\caption{\footnotesize Computational \& statistical performances of Newton, B-IPS and L-BFGS
 on general designs of large size.\label{tab:general}}

\begin{center}

\begin{tabular}{l cc cc cc}
\hline  & \multicolumn{2}{c}{$p=1,\negthinspace000$} & \multicolumn{2}{c}{$p=4,\negthinspace000$} & \multicolumn{2}{c}{$p={10,\negthinspace000}$}
\\
  & \multicolumn{2}{c}{$\epsilon_{\textrm{tol}}=1\mbox{e-}6$}  & \multicolumn{2}{c}{$\epsilon_{\textrm{tol}}=1\mbox{e-}6$} & \multicolumn{2}{c}{$t_{\max}=600$}
\\ 

\hline
& Time & estErr  & Time & estErr  & relGrad & estErr
\\
Newton   & $35.2$ & $0.12$ & $528.7$ & $0.11$ & -- & --
\\
B-IPS$^{1} $ & $16.8$ & $3.2$ & $116.4$  & $23.2$& $5.92$ & $109.3$
\\

\hline

\hline
& \multicolumn{2}{c}{$p=4,\negthinspace000$} & \multicolumn{2}{c}{$p={10,\negthinspace000}$}  & \multicolumn{2}{c}{$p={12,\negthinspace000}$}
\tabularnewline  & \multicolumn{2}{c}{$\epsilon_{\textrm{tol}}=1\mbox{e-}6$}  & \multicolumn{2}{c}{$t_{\max}=600$} & \multicolumn{2}{c}{$t_{\max}=600$}
\\

\hline   & Time & estErr  & relGrad & estErr  & relGrad & estErr
\\
L-BFGS &  $276.5$ & $2.6$ & $12250$ & $1022$ & -- & --
\\
B-IPS$^2$  & $94.2 $ & $2.8 $ & $5.6$ & $44.1$ & $12.8$ & $462.8$
\\
\hline
\end{tabular}
\end{center}
\end{table}

From the table, we  notice that B-IPS is less
precise than Newton in general, but having low computational complexity
makes it suitable for large-scale data applications where moderate accuracy
usually suffices. Indeed, Newton's method, when feasible,
gives the smallest estimation error, but it easily fails when $p$
is large (say $p\geq6000$). B-IPS$^{1}$ (with Newton as the sub-problem
solver and $g_{k}=200$) has better efficiency and scalability in
computation---see the case  when $p=10,\negthinspace000$, in particular.
The same conclusion can be drawn from the comparison between quasi-Newton
and B-IPS. In the experiments, L-BFGS
 ran out of memory when
$p>10,\negthinspace000$. B-IPS$^{2}$, which takes L-BFGS as the
sub-problem solver and  $g_{k}=2000$, showed   the best scalability.
In summary, the benefits brought by BCD, reparametrization, and  randomization
 are impressive in large-scale problems.
We  conducted even larger experiments (with $p\ge 20\mbox{,}000$) to study the scalability of B-IPS;  the reader may refer to  the Appendix
for more details.

\subsection{$\ell_{1}$-IPS on real data }

The dataset is collected from a Portuguese marketing campaign related
to bank deposit subscription \citep{moro2014}. We use $41,\negthinspace188$
instances and $10$ categorical variables to study whether a client
subscribes a term deposit or not. The information of these variables
is shown in Table \ref{tab:variables}. We group the data at each
observed combination level of all categorical variables, and use the
total number of successful subscriptions as the response variable.

\begin{table}[H]
\footnotesize
\begin{centering}
\begin{tabular}{l|l}
\hline
Variables & Description\\
\hline
\texttt{job} & type of job ($12$ levels)\\
\texttt{marital} & marital status ($4$ levels)\\
\texttt{education} & education level ($8$ levels)\\
\texttt{default} & whether the client has credit in default ($3$ levels)\\
\texttt{housing} & whether the client has housing loan ($3$ levels)\\
\texttt{loan} & whether the client has personal loan ($3$ levels)\\
\texttt{contact} & contact communication type, cellular or telephone ($2$ levels)\\
\texttt{month} & the month in which the last contact was made ($10$ levels)\\
\texttt{day} & the day of week when the last contact was made ($5$ levels)\\
\texttt{poutcome} & outcome of the previous marketing campaign ($3$ levels)\\
\hline
\end{tabular}
\par\end{centering}
\caption{\footnotesize Categorical variables   in the bank campaign data.\label{tab:variables}}
\end{table}

\begin{figure}[H]
\begin{centering}
\includegraphics[width=0.45\columnwidth]{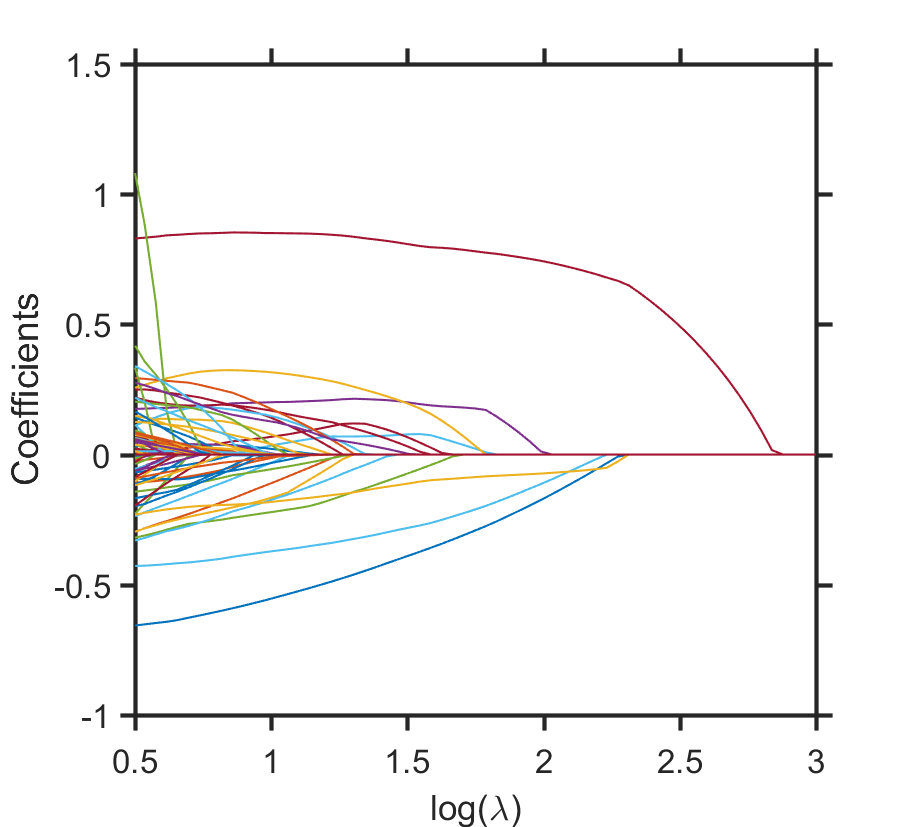}\includegraphics[width=0.45\columnwidth]{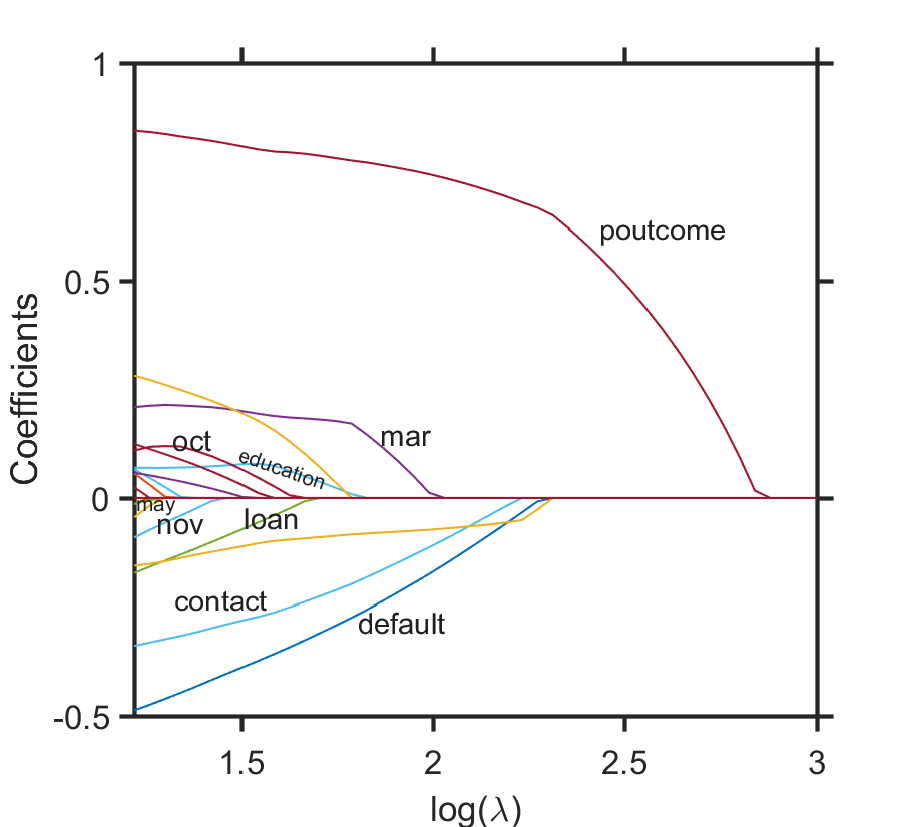}
\par\end{centering}

\centering{}\caption{\footnotesize Left: the $\ell_{1}$  solution path on the bank marketing data. Right:
the solution path including selected variables only. \label{fig:Solution path}}
\end{figure}

We consider a three-way association model, including all main effects,
second-order interactions, and third-order interactions. This results
in a large model with $5,\negthinspace874$ predictors. We ran $\ell_{1}$-IPS
to compute the solution path as shown in Figure \ref{fig:Solution path}
and used EBIC \citep{chen2008} for parameter tuning, where  a sparse model
with only $16$ predictors was obtained. See Table \ref{tab:variable selection} for the selected terms;  we also  labeled all main effects in Figure \ref{fig:Solution path}.
(We  ran the  experiment on a two-way association model, too,
and found that all the main terms and two-way interaction terms listed in
Table \ref{tab:variable selection} still got selected.)

Some interesting and useful  conclusions can be drawn from the variable selection
result. First,
\texttt{month} plays an important role in the marketing campaign, as supported by some previous studies (e.g., \citet{moro2014}).
In particular, clients are unlikely to subscribe a term deposit if
the last contact occurs in November. Moreover, contact communication
type being cellular and the outcome of previous marketing campaign
being successful are favorable factors for successful subscription.

According to the table, all of the selected two-way interactions involve the variable
\texttt{poutcome} = \texttt{non-exist} (no record in the previous
campaign), which separates out this special group of clients.  Our
model also contains two three-way terms, and notably they show no collinearity with the other selected variables, thereby worthy of consideration.
The positive coefficients of the two terms,  (\texttt{job} = \texttt{technician}) $*$ (\texttt{education} = \texttt{professional})$*$ (\texttt{marital} = \textrm{\texttt{married}}) and (\texttt{poutcome} = \texttt{non-exist}) $*$ (\texttt{education} = \texttt{univ}) $*$ (\texttt{marital} =  \textrm{\texttt{single}}), indicate that  the group of married  technicians who received
professional training  and the group of singles with university-level
degrees but no record in the previous marketing campaign  tend
to subscribe a term deposit.
\begin{table}[H]
{\footnotesize{
\noindent \begin{centering}
\begin{tabular}{l|ll}
\hline
\multirow{5}{*}{Main } & \texttt{month} = \texttt{mar} ($+$) & \texttt{month} = \texttt{may} ($-$)\\
 & \texttt{month} = \texttt{oct} ($+$) & \texttt{month} = \texttt{nov} ($-$)\\
 & \texttt{education} = \texttt{univ} ($+$) & \texttt{poutcome} = \texttt{success} ($+$)\\
 & \texttt{default} = \texttt{unkown} ($-$) & \texttt{contact} = \texttt{telephone} ($-$)\\
 & \texttt{loan} = \texttt{yes} ($-$) & \tabularnewline
\hline
\multirow{5}{*}{Two-way} & \multicolumn{2}{l}{(\texttt{poutcome} = \texttt{non-exist}) $*$ \texttt{month} = \texttt{mar}
($+$) }\\
 & \multicolumn{2}{l}{(\texttt{poutcome} = \texttt{non-exist}) $*$ \texttt{month} = \texttt{oct}
($+$)}\\
 & \multicolumn{2}{l}{(\texttt{poutcome} = \texttt{non-exist}) $*$ \texttt{education} = \texttt{univ}
($+$)}\\
 & \multicolumn{2}{l}{(\texttt{poutcome} = \texttt{non-exist}) $*$ \texttt{contact} = \texttt{telephone}
($-$)}\\
 & \multicolumn{2}{l}{(\texttt{poutcome} = \texttt{non-exist}) $*$ \texttt{housing} = \texttt{unkown}
($-$)}\\
\hline
\multirow{4}{*}{Three-way} & \multicolumn{2}{l}{(\texttt{job} = \texttt{technician}) $*$ (\texttt{education} = \texttt{professional}) }\\
 & \multicolumn{2}{l}{$*$ (\texttt{marital} = \textrm{\texttt{married}}) ($+$)}\\
 & \multicolumn{2}{l}{(\texttt{poutcome} = \texttt{non-exist}) $*$ \texttt{education} = \texttt{univ} }\\
 & \multicolumn{2}{l}{$*$ (\texttt{marital} =  \textrm{\texttt{single}}) ($+$)}\\
\hline
\end{tabular}
\par\end{centering}
}}
\caption{\footnotesize Selected variables with signs of coefficients in parentheses.\label{tab:variable selection}}
\end{table}

\newpage
\appendix

\section{Appendix}

\subsection{Proof of Theorem \ref{prop:sequence convergence CD}}

The result is an application of Theorem 2.1 of \citet*{Luo1992}.
In fact, under the assumptions made in Section \ref{subsec:model}
and the assumption in this theorem, $g(\boldsymbol{X}\boldsymbol{\beta})\triangleq\boldsymbol{q}^{T}\exp(\boldsymbol{X}\boldsymbol{\beta})$
is strictly convex and twice continuously differentiable on its effective
domain, and $\nabla^{2}g(\boldsymbol{X}\hat{\boldsymbol{\beta}})$
is positive definite. It follows from Theorem 2.1 of \citet*{Luo1992}
that $\{\boldsymbol{\beta}^{(t)}\}_{t=0}^{\infty}$ generated by the
Algorithm \ref{alg:CD alg} (with cyclic update) converges to $\hat{\boldsymbol{\beta}}$
at least linearly.
\subsection{Proof of Theorem \ref{prop:IIS}}

We prove the conclusion by induction. Given $\mathring{\boldsymbol{\beta}}^{(t)}$,
$\mathring{\boldsymbol{\mu}}^{(t)}$, $\bar{\boldsymbol{\mu}}^{(t)}$
with $\bar{\boldsymbol{\mu}}^{(t)}=\mathring{\boldsymbol{\mu}}^{(t)}/\langle\boldsymbol{1},\mathring{\boldsymbol{\mu}}^{(t)}\rangle$,
it suffices to show that (i) equation (\ref{eq:log-concave update beta})
and equation (\ref{eq:IIS_beta}) yield the same solution, and (ii)
$\bar{\boldsymbol{\mu}}^{(t+1)}$ obtained from (\ref{eq:IIS_mu})
and $\mathring{\boldsymbol{\mu}}^{(t+1)}$ updated from (\ref{eq:log-concave update mu})
satisfy $\bar{\boldsymbol{\mu}}^{(t+1)}=\mathring{\boldsymbol{\mu}}^{(t+1)}/\langle\boldsymbol{1},\mathring{\boldsymbol{\mu}}^{(t+1)}\rangle$.

Plugging $\bar{\boldsymbol{\mu}}^{(t)}=\mathring{\boldsymbol{\mu}}^{(t)}/\langle\boldsymbol{1},\mathring{\boldsymbol{\mu}}^{(t)}\rangle$
into (\ref{eq:log-concave update beta}) gives:

\[
\langle\boldsymbol{1},\boldsymbol{n}\rangle\sum_{i}\mathring{x}_{ij}\bar{\mu}_{i}^{(t)}\exp[\mathring{x}_{i+}(\mathring{\beta}_{j}^{(t+1)}-\mathring{\beta}_{j}^{(t)})]=\sum_{i}n_{i}\mathring{x}_{ij},
\]
or equivalently

\[
\sum_{i}\mathring{x}_{ij}\bar{\mu}_{i}^{(t)}\exp[\mathring{x}_{i+}(\mathring{\beta}_{j}^{(t+1)}-\mathring{\beta}_{j}^{(t)})]=\sum_{i}(n_{i}/\langle\boldsymbol{1},\boldsymbol{n}\rangle)\mathring{x}_{ij},
\]
which is exactly (\ref{eq:IIS_beta}).

Next, given $\mathring{\boldsymbol{\beta}}^{(t+1)}$, we have
\begin{eqnarray*}
\frac{\mathring{\boldsymbol{\mu}}^{(t+1)}}{\langle\boldsymbol{1},\mathring{\boldsymbol{\mu}}^{(t+1)}\rangle} & = & \frac{\mathring{\boldsymbol{\mu}}^{(t)}\circ\exp[\mathring{\boldsymbol{X}}(\mathring{\boldsymbol{\beta}}^{(t+1)}-\mathring{\boldsymbol{\beta}}^{(t)})]}{\langle\boldsymbol{1},\mathring{\boldsymbol{\mu}}^{(t)}\circ\exp[\mathring{\boldsymbol{X}}(\mathring{\boldsymbol{\beta}}^{(t+1)}-\mathring{\boldsymbol{\beta}}^{(t)})]\rangle}\\
 & = & \frac{[\mathring{\boldsymbol{\mu}}^{(t)}/\langle\boldsymbol{1},\mathring{\boldsymbol{\mu}}^{(t)}\rangle]\circ\exp[\mathring{\boldsymbol{X}}(\mathring{\boldsymbol{\beta}}^{(t+1)}-\mathring{\boldsymbol{\beta}}^{(t)})]}{\langle\boldsymbol{1},[\mathring{\boldsymbol{\mu}}^{(t)}/\langle\boldsymbol{1},\mathring{\boldsymbol{\mu}}^{(t)}\rangle]\circ\exp[\mathring{\boldsymbol{X}}(\mathring{\boldsymbol{\beta}}^{(t+1)}-\mathring{\boldsymbol{\beta}}^{(t)})]\rangle}\\
 & = & \frac{\bar{\boldsymbol{\mu}}^{(t)}\circ\exp[\mathring{\boldsymbol{X}}(\mathring{\boldsymbol{\beta}}^{(t+1)}-\mathring{\boldsymbol{\beta}}^{(t)})]}{\langle\boldsymbol{1},\bar{\boldsymbol{\mu}}^{(t)}\circ\exp[\mathring{\boldsymbol{X}}(\mathring{\boldsymbol{\beta}}^{(t+1)}-\mathring{\boldsymbol{\beta}}^{(t)})]\rangle}\\
 & = & \bar{\boldsymbol{\mu}}^{(t+1)},
\end{eqnarray*}
where 
the third equality  follows from the induction hypothesis, and the last equality is due
to (\ref{eq:IIS_mu}).

\subsection{Proof of Theorem \ref{lem:l1 univariate}}

Let $f_{1}(\beta)=-\langle\boldsymbol{n},\boldsymbol{x}\rangle\beta+\langle\boldsymbol{q},\exp(\beta\boldsymbol{x})\rangle+\lambda\lvert\beta\rvert$.
From the Karush\textendash Kuhn\textendash Tucker equation
\begin{equation}
\boldsymbol{x}^{T}\boldsymbol{\mu}-\boldsymbol{x}^{T}\boldsymbol{n}+\lambda s(\beta)=0,\label{eq:KKT uni-1}
\end{equation}
where $\boldsymbol{\mu}=\boldsymbol{q}\circ\exp(\beta\boldsymbol{x})$
and the sub-gradient $s(\beta)$ satisfies $s(\beta)=\sgn(\beta)$
if $\beta\neq0$ and $s(\beta)\in[-1,1]$ if $\beta=0$. For such
a convex problem, this equation is necessary and sufficient for the
solution.

Recall (\ref{eq:binary}), and so we have $\boldsymbol{x}^{T}\boldsymbol{\mu}=\langle\boldsymbol{x},\boldsymbol{q}\rangle\exp\beta$.
Plugging this into (\ref{eq:KKT uni-1}) gives $\langle\boldsymbol{x},\boldsymbol{q}\rangle\exp\beta-\boldsymbol{x}^{T}\boldsymbol{n}+\lambda s(\beta)=0$.
The solution thus follows

\[
\hat{\beta}=\begin{cases}
\log \frac{\langle\boldsymbol{x},\boldsymbol{n}\rangle-\lambda }{\langle\boldsymbol{x},\boldsymbol{q}\rangle}, & \mbox{ if } \langle\boldsymbol{x},\boldsymbol{n}-\boldsymbol{q}\rangle\geq\lambda\\
0, & \mbox{ if } -\lambda<\langle\boldsymbol{x},\boldsymbol{n}-\boldsymbol{q}\rangle<\lambda\\
\log \frac{\langle\boldsymbol{x},\boldsymbol{n}\rangle+\lambda}{\langle\boldsymbol{x},\boldsymbol{q}\rangle}, & \mbox{ if }\langle\boldsymbol{x},\boldsymbol{n}-\boldsymbol{q}\rangle\leq-\lambda.
\end{cases}
\]

In the $\ell_0$ case, let $f_{0}(\beta)=-\langle\boldsymbol{n},\boldsymbol{x}\rangle\beta+\langle\boldsymbol{q},\exp(\beta\boldsymbol{x})\rangle+\lambda1_{\beta\neq0}$
which is nonconvex, and $\hat{\beta}$ a global minimizer of $f_{0}$. Notice that given
a binary vector $\boldsymbol{x}$, we have $$\exp(\beta\boldsymbol{x})=\exp(\beta)\boldsymbol{x}+(\boldsymbol{1}-\boldsymbol{x}),$$
from which it follows that
\begin{equation}
f_{0}(\beta)=-\langle\boldsymbol{n},\boldsymbol{x}\rangle\beta+\langle\boldsymbol{q},\boldsymbol{x}\rangle\exp(\beta)+\langle\boldsymbol{q},\boldsymbol{1}-\boldsymbol{x}\rangle+\lambda1_{\beta\neq0}.\label{eq:f0 binary}
\end{equation}
If $\hat{\beta}\neq0$,  $\partial f_{0}/\partial\beta=0$
gives $\hat{\beta}=\log(\langle\boldsymbol{n},\boldsymbol{x}\rangle/\langle\boldsymbol{q},\boldsymbol{x}\rangle)$,
where $\langle\boldsymbol{n},\boldsymbol{x}\rangle\neq\langle\boldsymbol{q},\boldsymbol{x}\rangle$, and so
$
f_{0}(\hat{\beta})=-\langle\boldsymbol{n},\boldsymbol{x}\rangle\log(\langle\boldsymbol{n},\boldsymbol{x}\rangle/\langle\boldsymbol{q},\boldsymbol{x}\rangle)+\langle\boldsymbol{n},\boldsymbol{x}\rangle+\langle\boldsymbol{q},\boldsymbol{1}-\boldsymbol{x}\rangle+\lambda.
$

The condition $f_{0}(\hat{\beta})\le f_{0}(0)=\langle\boldsymbol{q},\boldsymbol{1}\rangle$
can be expressed as
\[
-\langle\boldsymbol{n},\boldsymbol{x}\rangle\log(\langle\boldsymbol{n},\boldsymbol{x}\rangle/\langle\boldsymbol{q},\boldsymbol{x}\rangle)+\langle\boldsymbol{n},\boldsymbol{x}\rangle-\langle\boldsymbol{q},\boldsymbol{x}\rangle+\lambda\le 0,
\]
which is equivalent to $\DKL(\langle\boldsymbol{n},\boldsymbol{x}\rangle\|\langle\boldsymbol{q},\boldsymbol{x}\rangle)\ge \lambda$.
In summary, a globally optimal solution of $\min_{\beta\in\mathbb{R}}f_{0}(\beta)$
is
\[
\hat{\beta}=\begin{cases}
0, & \textrm{if}\quad\mathbf{D}_{\mbox{\tiny KL}}(\langle\boldsymbol{n},\boldsymbol{x}\rangle\|\langle\boldsymbol{q},\boldsymbol{x}\rangle)< \lambda,\\
\log(\langle\boldsymbol{n},\boldsymbol{x}\rangle/\langle\boldsymbol{q},\boldsymbol{x}\rangle), & \textrm{if}\quad\mathbf{D}_{\mbox{\tiny KL}}(\langle\boldsymbol{n},\boldsymbol{x}\rangle\|\langle\boldsymbol{q},\boldsymbol{x}\rangle)\ge \lambda.
\end{cases}
\]
It is worth mentioning that $\hat{\beta}$ is not unique if $\DKL(\langle\boldsymbol{n},\boldsymbol{x}\rangle\|\langle\boldsymbol{q},\boldsymbol{x}\rangle)=\lambda$,
in which case $\hat\beta$ can  take either $0$ or $\log(\langle\boldsymbol{n},\boldsymbol{x}\rangle/\langle\boldsymbol{q},\boldsymbol{x}\rangle)$.

\subsection{Large experiments}
This part  tests the performance of B-IPS on large tables and large designs  with dimensionality $p\ge 20000$. The experiments were performed  on a  machine    with  2.9GHz CPU and  64GB RAM installed. We consider  6 examples.
The first three   come from  contingency tables,  consisting of  $200$, $225$ and $250$ binary categorical variables, respectively. All
interactions (up to second order) are included, resulting in   $p=20,\negthinspace101$, $p=25,\negthinspace426$, and  $p=31,\negthinspace376$, respectively. The observed entries are sampled from the table and the number $N$ is fixed   at $100,\negthinspace000$.
In generating the  coefficients, we let the intercept take $5$,   and set all  $\beta_j^*$ to  zero except the last $100$  which  are sampled from
$\mathcal{N}(-1,1)$. The error tolerance $\epsilon_{\textrm{tol}}$ in the stopping criterion is $ 1\mbox{e-}5$. The remaining  three examples have  large  design matrices involving non-binary features, where   $N=100,\negthinspace000$  and  $p$ varies from $20,\negthinspace000$
to $30,\negthinspace000$. In these examples,  $\beta_{0}^{*}=10$ and $\beta_{j}^{*}$ are i.i.d. following $ 0.5\mathcal{N}(10,1)+0.5\mathcal{N}(-10,1)$,
 $1\leq j\leq p-1$. The prototype design matrix is scaled,  $\mathring{\boldsymbol{X}}\leftarrow\mathring{\boldsymbol{X}}/(200\lVert\mathring{\boldsymbol{X}}\rVert_{\max})$, and   $\epsilon_{\textrm{tol}}=1\mbox{e-}7$. The block sizes in calling B-IPS are fixed at 5000 (or approximately so).
The computational time and estimation error for
 are reported in the following table.

\begin{table}[h!]
\setlength{\tabcolsep}{3pt}     
\renewcommand{\arraystretch}{1} 
{\footnotesize{
\caption{\footnotesize Computational and statistical performance of B-IPS on large tables and designs, where  Time  is in seconds, and Err denotes the relative estimation error defined in Section \ref{sec:exp}. \label{tab:large_table}}
\begin{center}
\begin{tabular}{l cc cc cc cc cc cc}

\hline
  & \multicolumn{2}{c}{ Ex $1$} &   \multicolumn{2}{c}{Ex $2$}   & \multicolumn{2}{c}{ Ex $3$ } & \multicolumn{2}{c}{ Ex $4$} &   \multicolumn{2}{c}{Ex $5$}   & \multicolumn{2}{c}{ Ex $6$ }
\\  & \multicolumn{2}{c}{$p=20,\negthinspace101$} & \multicolumn{2}{c}{$p=25,\negthinspace426$} & \multicolumn{2}{c}{$p=31,\negthinspace376$} & \multicolumn{2}{c}{$p=20,\negthinspace000$} & \multicolumn{2}{c}{$p=25,\negthinspace000$} & \multicolumn{2}{c}{$p={30,\negthinspace000}$}
\\

\hline & Time & Err  & Time & Err & Time & Err & Time & Err  & Time & Err  & Time & Err
\\
& 2.9e+4  &$0.039$   & 3.5e+4   &$0.24$   & 3.9e+4   &$0.33$ & 1.8e+3 & $0.0005$ &  2.8e+3  & $0.002$&  4.2e+3 & $0.005$
\\
\hline
\end{tabular}
\end{center}
}}
\end{table}

\bibliographystyle{apalike}
\bibliography{IPSBib}

\end{document}